# Polar molecular ordering in the N$_X$ phase of bimesogens and enantiotopic discrimination in the NMR spectra of rigid prochiral solutes.

*Anant Kumar, Alexandros G. Vanakaras[1] and Demetri J. Photinos*

Department of Materials Science, University of Patras, Patras 26504, Greece

**Abstract:** The potential of mean torque governing the orientational ordering of prochiral solutes in the two nematic phases (N and N$_X$) formed by certain classes of symmetric achiral bimesogens is formulated and used for the analysis of existing NMR measurements on solutes of various symmetries dissolved in the two phases. Three distinct attributes of the solvent phase, namely polarity of the orientational ordering, chirality of the constituent molecules and spatial modulation of the local director, are identified as underlying three possible mechanisms for the generation of chiral asymmetry in the low temperature nematic phase (N$_X$). The role and quantitative contribution of each mechanism to enantiotopic discrimination in the N$_X$ phase are presented and compared with the case of the conventional chiral nematic phase (N*). It is found that polar ordering is essential for the appearance of enantiotopic discrimination in small rigid solutes dissolved in the N$_X$ phase and that such discrimination is restricted to solutes belonging to the point group symmetries $C_s$ and $C_{2v}$.

---

[1] Email: a.g.vanakaras@upatras.gr

# 1. Introduction.

One of the most fascinating recent advances in liquid crystal science is the discovery of a second, lower temperature, nematic phase formed by certain types of bimesogens.[1–5] This phase, to be referred to as $N_X$ in the present work, although optically uniaxial, shows distinct differences from the conventional (uniaxial, a-chiral and a-polar) nematic phase (denoted here by N), notably in its appearance in polarized microscopy, its response to electric[6] and magnetic[7] fields, its X-ray diffraction,[8,9] dielectric[10] and NMR spectroscopy.[11–13] A clear distinguishing feature of the $N_X$ phase is the doubling of dipolar and quadrupolar NMR spectral lines associated with pairs of prochiral molecular segments which give coincident spectral lines in the N phase. This doubling has first been observed[14] in prochiral solute molecules dissolved in chiral nematic (N*) solvents and is referred to as enantiotopic discrimination[15,16]; it indicates the loss of equivalence in the orientational ordering of the members of a prochiral pair of segments on going from an achiral to a chiral nematic environment. Thus, the appearance of enantiotopic discrimination in the NMR spectra of the $N_X$ phase of neat dimers, or of probe-solutes dissolved therein, has been taken as an indication of the onset of some sort of chiral orientational ordering across the N to $N_X$ phase transition. Given that the $N_X$ phase can be formed by achiral dimer molecules, such chiral ordering could in principle result from the chiral stacking of the flexible dimer molecules; in turn, these molecules would be subjected to a chiral biasing of their conformational statistics and thus lose their strict statistical achirality, as conformational enantiomers would no longer come with identical probabilities. Information on the possible origin of such chiral stacking of the molecules came from independent experimental observations indicating the helical twisting of the directions of preferential local ordering in the $N_X$ phase.[17,18,13]

In an explicit molecular model proposed for the $N_X$ phase,[13] the N-$N_X$ phase transition is described as a transition from an apolar phase to a polar one via spontaneous symmetry breaking. As a result of this mechanism,[19] the macroscopic apolarity is preserved through the twisting of the direction of polar ordering (the polar director), perpendicular to a fixed direction (the axis of helical twisting), while the overall achirality is manifested through the equiprobable appearance of right-handed and left-handed twisted domains. The helical twisting of the polar director forces the principal axis of the ordering tensor of the mesogenic units to also twist, albeit forming a fixed angle with the helix axis (the "tilt" angle).

Although a byproduct of the polar director helical twisting, the presence of this "tilted twisting" has led some authors to identifying the $N_X$ phase with the twist-bend nematic phase ($N_{TB}$) originally proposed by R.B. Meyer.[20] However, the original proposal referred (i) to length scales of the pitch that are larger than the typical $N_X$ pitch lengths by two orders of magnitude and (ii) to twisting & bending of the nematic director, i.e. of a local axis of full rotational symmetry, a situation which clearly does not apply to the $N_X$ phase.[19] Accordingly, we feel that the identification of the $N_X$ phase as a twist-bend nematic, although widespread, is not in agreement with the $N_{TB}$ phase originally proposed by R.B. Meyer, and we will continue to refer to it in this work by the symbol ($N_X$) that has been used for it in the original literature where its experimental discovery and identification were first reported.[2,6]

Here, based on the symmetry of the environment sensed by a rigid solute molecule in the $N_X$ phase, derived from the molecular model in ref. [19], we obtain the general form of the potential of mean torque that governs the orientational distribution of such molecule in the $N_X$ phase, which in turn determines the structure of the respective NMR spectra. We demonstrate that, unlike the potential of mean torque in the $N^*$ phase[21], the polar terms are essential for the enantiotopic discrimination while the terms associated with the molecular chirality of the solute, which underlie enantiotopic discrimination in the $N^*$ phase, are less significant for small solute molecules in the $N_X$ phase. We also show that the point symmetry requirements on the solute molecules for the appearance of the primary phenomenon are different in the $N_X$ phase.

## 2. $N_X$ phase symmetry and solute potential of mean torque.

As detailed in ref. [19], the $N_X$ phase consists of domains. Each domain is locally polar, with the direction of polar ordering designated as the director $\hat{m}$. Within each domain, this director undergoes constant-pitch helical twisting about a macroscopic axis whose direction remains perpendicular to $\hat{m}$ and is denoted by $\hat{n}_h$. Accordingly, the local polarity is effectively averaged out over the length scale of the helical domain. A third axis $\hat{l}_h$, perpendicular to both $\hat{n}_h$ and $\hat{m}$, and therefore following the helical twisting of $\hat{m}$, is introduced to provide an orthogonal local frame of axes for each domain. It is important to note that the presence of only one local director, $\hat{m}$, implies that any tensor quantity will have this local director as a principal axis but its other two principal axes, necessarily lying on the $\hat{n}_h$-$\hat{l}_h$ plane, will in general deviate from the axes $\hat{n}_h$ and $\hat{l}_h$. This deviation has been termed[19] as "eccentricity" of the ordering, when viewed in the direction of the helix axis $\hat{n}_h$, or as "tilt" when viewed in the direction of $\hat{m}$, i.e. on the $\hat{n}_h$-$\hat{l}_h$ plane, and is reversed on reversing the twisting sense of $\hat{m}$ about $\hat{n}_h$.

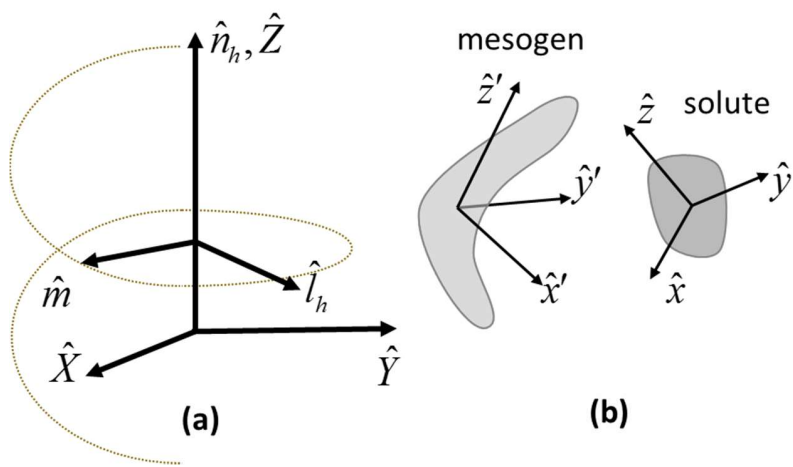

**Figure 1.** (a) Illustration of the twisting of the polar director $\hat{m}$ in a domain of the $N_X$ phase. Also shown are the macroscopic axes $X, Y, Z$ of the domain and the axes $\hat{m}, \hat{l}_h, \hat{n}_h$ in which the local orientational ordering of the molecules in the domain is described. (b) Molecular axes fixed on the mesogenic (solvent) and on the solute molecules.

An unbiased sample has statistically equal populations of domains with right- and left-handed helical twisting. Thus, such sample is overall apolar and achiral, despite the locally strong polar ordering and the handedness of the individual domains which results from the tight twisting of the polar director $\hat{m}$, see Figure 1(a). Furthermore, the sample will appear macroscopically uniaxial if a uniform alignment of the helical axes $\hat{n}_h$ of the domains is imposed by some aligning stimulus. Therefore, the experimental probing of polar ordering and of chiral structures would require techniques which are sensitive to the local ordering of the molecules. NMR offers precisely such techniques, wherein the measured spectra from a probe molecule are related to the orientational ordering imposed on that molecule by its local environment. Such ordering can be derived from the potential of mean torque which embodies the interaction of the probe molecule with its local environment and thereby governs its probability distribution. The local symmetry within each domain of the $N_X$ phase is a twofold rotation about the $\hat{m}$ axis,[19] which implies invariance with respect to the simultaneous inversion of the other two axes: $(\hat{n}_h, \hat{l}_h) \Leftrightarrow (-\hat{n}_h, -\hat{l}_h)$. The helical twisting implies the spatial variation of the directions $\hat{m}$ and $\hat{l}_h$ as follows

$$\begin{aligned}\hat{m}(Z) &= -\hat{X}\sin k(Z+Z_0) + \hat{Y}\cos k(Z+Z_0) \\ \hat{l}_h(Z) &= \hat{X}\cos k(Z+Z_0) + \hat{Y}\sin k(Z+Z_0)\end{aligned} \quad , \qquad (1)$$

where $Z$ is the coordinate along a macroscopic axis $\hat{Z}$ whose direction coincides with the helix axis $\hat{n}_h$, the macroscopic axes $\hat{X}$ and $\hat{Y}$ are mutually perpendicular in the plane normal to $\hat{n}_h$, the helix constant $k$ is related to the helical pitch $h$ by $k = 2\pi/h$ and $Z_0$ is a phase factor that can vary from one domain to another.

**For a solute molecule of negligible spatial extension**, its instantaneous environment is a region of molecular dimensions within a helical domain. We therefore formulate the potential of mean torque within such a domain, subject to the local symmetries of the domain and the molecular symmetries of the solute. To this end, assuming that the dimensions of the solute molecule are small compared to the pitch of the $N_X$ helix, the "center" of the molecule can be taken as the effective position of the entire molecule and all of its parts are considered to experience the environment which corresponds to that position. Since the phase shows no variation of the orientational ordering on moving along the $X$ or $Y$ directions, only the $Z$ coordinate of the molecular center is relevant. The potential of mean torque, $V(\omega; Z)$, experienced by such a rigid solute molecule is related to the probability density, $f(\omega; Z)$, of finding the solute molecule in a given orientation $\omega$ when the solvent is positioned at $Z$, according to

$$f(\omega; Z) = \zeta^{-1} \exp[-V(\omega; Z)] \quad , \qquad (2)$$

with the usual normalization factor $\zeta = \int \exp[-V(\omega; Z)]d\omega$ being independent of the $Z$-coordinate.

To determine the leading tensor terms of $V(\omega; Z)$ we follow a formulation analogous to the one presented in [21] for the chiral nematic phase, N*, but here adapted to the symmetries of the conventional nematic phase, N, and of the $N_X$ phase. Accordingly, the molecular frame axes, see Figure 1(b), will be denoted generically by the unit vectors $\hat{a}, \hat{b} = \hat{x}, \hat{y}, \hat{z}$ (for the solute) and $\hat{a}', \hat{b}' = \hat{x}', \hat{y}', \hat{z}'$ (for the solvent) and the local phase axes by $\hat{I}, \hat{J} = \hat{n}_h, \hat{l}_h, \hat{m}$.

The lowest (first) rank terms of $V(\omega; Z)$ are of the "vector-vector" form

$$V^{(1)}(\omega;Z) = \overline{C}^{(1)}_{a,a'} \langle \hat{a} \cdot \hat{a}' \rangle' = \overline{C}^{(1)}_{a,a'} \langle (I \cdot \hat{a}') \rangle' (\hat{a} \cdot \hat{I}) \quad , \tag{3}$$

where the primed angular brackets indicate averaging of the solvent molecule conformations and orientations relative to the phase-fixed axes; summation over repeated tensor indices/axis labels is implied throughout. The quantities $\overline{C}^{(1)}_{a,a'}$ are effective molecular coupling parameters, understood to be obtained after averaging the vector part of the solvent-solute interaction over the solvent molecule positions. All the terms in eq (3) vanish for the N phase, due to its complete apolarity, whereas in the $N_X$ phase, only the terms involving the polar director $\hat{m}$ can survive, provided the solute molecule has at least one polar axis that couples to that director. Therefore, one generally has, for the first rank contribution in the $N_X$ phase,

$$V^{(1)}(\omega;Z) = \langle g^{(1)}_a \rangle' (\hat{a} \cdot \hat{m}) \quad , \tag{4}$$

where the effective coupling coefficients, defined as $\langle g^{(1)}_a \rangle' = \overline{C}^{(1)}_{a,a'} \langle (\hat{m} \cdot \hat{a}') \rangle'$, vanish by symmetry in case the solute molecule is apolar with respect to the molecular axis $a$. The implicit $Z$-dependence of $V^{(1)}$ in eq (4) is due to the helical twisting of $\hat{m}$, as described in eq (1).

The symmetric second rank contribution is

$$V^{(2)}(\omega;Z) = \overline{C}^{(2)}_{ab;a'b'} \left( \frac{3}{2} \langle (\hat{a} \cdot \hat{a}')(\hat{b} \cdot \hat{b}') \rangle' - \frac{1}{2} \delta_{ab} \delta_{a'b'} \right) \quad . \tag{5}$$

Taking into account the symmetry of the $N_X$ phase, this contribution can be put in the form

$$V^{(2)}(\omega;Z) = \langle g^{(2)}_{ab} \rangle' S^{n_h}_{ab} + \langle \Delta^{(2)}_{ab} \rangle' \left( S^{l_h}_{ab} - S^m_{ab} \right) + \langle \Phi^{(2)}_{ab} \rangle' \left( q^{n_h l_h}_{ab} + q^{n_h l_h}_{ba} \right) \quad , \tag{6}$$

where the orientational second rank tensor components $a,b$ for the various local phase axes $I, J$ are defined as follows:

$$S^I_{ab} \equiv \frac{3}{2}(\hat{a} \cdot \hat{I})(\hat{b} \cdot \hat{I}) - \frac{1}{2}\delta_{ab} \quad ; \quad q^{IJ}_{ab} \equiv (\hat{a} \cdot \hat{I})(\hat{b} \cdot \hat{J}) \quad ; \quad \hat{I}, \hat{J} = \hat{n}_h, \hat{l}_h, \hat{m} \quad , \tag{7}$$

and the coupling coefficients to the solvent ordering, $\langle g^{(2)}_{ab} \rangle'$ (uniaxial contribution), $\langle \Delta^{(2)}_{ab} \rangle'$ (biaxial contribution) and $\langle \Phi^{(2)}_{ab} \rangle'$ (contribution of the "eccentricity" of the ordering about the polar director $\hat{m}$, or, equivalently, of the "tilt" of the other two principal axes of that ordering in the $\hat{n}_h - \hat{l}_h$ plane), are defined in terms of the molecular coupling parameters $\overline{C}^{(2)}_{ab;a'b'}$ and the solvent order parameters as follows:

$$\begin{aligned}\left\langle g^{(2)}_{ab}\right\rangle' &\equiv \overline{C}^{(2)}_{ab;a'b'}\left\langle S^{n_h}_{a'b'}\right\rangle' \\ \left\langle \Delta^{(2)}_{ab}\right\rangle' &\equiv \frac{1}{3}\overline{C}^{(2)}_{ab;a'b'}\left\langle S^{l_h}_{a'b'} - S^{m}_{a'b'}\right\rangle' \\ \left\langle \Phi^{(2)}_{ab}\right\rangle' &\equiv \overline{C}^{(2)}_{ab;a'b'}\frac{3}{2}\left\langle (\hat{n}_h\cdot\hat{a}')(\hat{l}_h\cdot\hat{b}')\right\rangle'\end{aligned} \qquad , \qquad (8)$$

In the uniaxial apolar nematic phase N, all the $\left\langle \Delta^{(2)}_{ab}\right\rangle'$ and the $\left\langle \Phi^{(2)}_{ab}\right\rangle'$ vanish by symmetry and the helix axis direction $\hat{n}_h$ is replaced by the usual nematic director $\hat{n}$. In the $N_X$ phase, the coupling coefficients $\left\langle \Phi^{(2)}_{ab}\right\rangle'$ invert their sign on inverting the sense of twisting of the $\hat{m}$ director whilst the $\left\langle g^{(2)}_{ab}\right\rangle'$ and $\left\langle \Delta^{(2)}_{ab}\right\rangle'$ coefficients remain invariant. The simultaneous presence of polar ordering and twisting of the polar director confers to the twisted domains of the $N_X$ phase a "structural chirality" which underlies many of the characteristic features of the phase that is formed by achiral molecules.

As shown in reference [21], in addition to the above leading rank terms, vector-pseudovector couplings could contribute to the solute potential of mean torque if the solvent molecules are chiral. On the other hand, it is known that the $N_X$ phase is obtained from strictly achiral (in the statistical sense) dimer molecules. However, the statistical achirality of the dimers is shown theoretically[19] to be preserved only in the N phase and to be broken in the $N_X$ phase, because of a small imbalance in the populations of conformers with opposite handedness. Therefore some weak contribution from vector-pseudovector couplings could be expected in the $N_X$ phase of statistically achiral dimers. This molecular chirality is quite distinct from the structural chirality defined in the previous paragraph, although in the case of flexible achiral dimers the molecular chirality may be thought as being induced by the structural chirality. Naturally, no molecular chirality would be obtained in an $N_X$ phase formed by strictly rigid achiral molecules.

The inclusion of the vector-pseudovector couplings, based on the assumption that the solvent molecules do not remain strictly achiral in the $N_X$ phase, would introduce the following additional contribution to the solute potential of mean torque[21]:

$$V^*(\omega, Z) = \overline{C}^*_{a \neq b, c; a' \neq b', c'} \left[ \left\langle \left( (\hat{a} \times \hat{b}) \cdot \hat{c}' \right) \left( (\hat{a}' \times \hat{b}') \cdot \hat{c} \right) \right\rangle' - \frac{1}{3} \varepsilon_{abc} \varepsilon_{a'b'c'} \right] . \qquad (9)$$

Here the coefficients $\overline{C}^*_{a \neq b, c; a' \neq b', c'}$ reflect the handedness of the solvent molecules, being of opposite signs for molecules of opposite handedness. Accordingly, mirror image chiral domains of the $N_X$ phase correspond to opposite signs for each of these coefficients. A number of surviving terms, dictated by the symmetries of the $N_X$ phase and depending on the molecular symmetries of the solute, is possible in this part of the potential of mean torque. These are given explicitly in section 5 and in Appendix II for rigid solute molecules of $C_{2v}$ point group symmetry.

Collecting the terms from eqs (4), (6) and (9) we have the following composition for the leading tensor contributions to the potential of mean torque experienced by a small rigid solute molecule in the $N_X$ phase formed by statistically achiral flexible dimer molecules:

$$V(\omega, Z) = V^{(1)}(\omega, Z) + V^{(2)}(\omega, Z) + V^*(\omega, Z) . \qquad (10)$$

The generalization to the case of solute molecules of non-negligible spatial extension is presented in Appendix I.

## 3. NMR spectral lines, solute order parameters and molecular structure.

In this work we shall consider NMR spectra obtained from residual proton dipolar couplings and deuterium quadrupolar couplings. For both types of couplings, the general spectral signature[21] of a site $s$, belonging to a solute molecule that is found at a fixed $Z$ coordinate within a domain of the $N_X$ phase and undergoes rapid (in the NMR measurement time scale) reorientational motion, is given by:

$$\nu_s / \nu_s^0 = \left\langle P_2(\hat{H} \cdot \hat{e}_s) \right\rangle , \qquad (11)$$

where $P_2(x) = (3x^2 - 1)/2$ denotes the second Legendre polynomial, $\hat{e}_s$ is the direction of the coupling (inter-proton vector or principal axis of the electric field gradient tensor at the position of the quadrupole moment of the deuterium nucleus*), $\hat{H}$ is the direction of the spectrometer magnetic field, the angular brackets indicate thermal averaging over the orientations of the solute molecule and $v_s^0$ are site-specific constants.[1]

The presently available NMR measurements on rigid solutes in the $N_X$ phase are for magnetically aligned samples, wherein the helix axis direction $n_h$ is parallel to the spectrometer magnetic field. In such case, the respective spectral signatures $(v_s/v_s^0)_\parallel$ are expressed in terms of the solute orientational order parameters $\langle S_{ab}^{n_h} \rangle$ as follows:

$$(v_s/v_s^0)_\parallel \equiv \langle P_2(\hat{n}_h \cdot \hat{e}_s) \rangle = S_{zz}^s \langle S_{zz}^{n_h} \rangle + \frac{1}{3}(S_{xx}^s - S_{yy}^s)(\langle S_{xx}^{n_h} \rangle - \langle S_{yy}^{n_h} \rangle)$$
$$+ \frac{4}{3}\left(S_{xy}^s \langle S_{xy}^{n_h} \rangle + S_{xz}^s \langle S_{xz}^{n_h} \rangle + S_{yz}^s \langle S_{yz}^{n_h} \rangle \right) \quad (12)$$

with the site geometrical constants defined as

$$S_{ab}^s = \frac{3}{2}(\hat{a}\cdot\hat{e}_s)(\hat{b}\cdot\hat{e}_s) - \frac{1}{2}\delta_{ab} \quad (13)$$

It should be noted in eq (13) that the spectral signatures $(v_s/v_s^0)_\parallel$ do not depend on the $Z$ coordinate within the domain.

In anticipation of future measurements on rigid solutes with the $N_X$ helix direction $n_h$ held perpendicular to the magnetic field, we give here the expression for the respective spectral signatures $(v_s/v_s^0)_\perp$. Taking the magnetic field in eq(11) to be along the macroscopic $\hat{Y}$ axis of eq. (1) and making use of the twofold rotational symmetry about the director $\hat{m}$, we have:

$$(v_s/v_s^0)_\perp = -\frac{1}{2}(v_s/v_s^0)_\parallel + \frac{1}{2}\langle P_2(\hat{m}\cdot\hat{e}_s) - P_2(\hat{l}_h\cdot\hat{e}_s)\rangle\cos 2kZ \quad . \quad (14)$$

---

[1] For quadrupolar couplings, eq (11) has an additional contribution associated with the biaxiality of the electric field gradient tensor, $\frac{\eta_{EFG}^s}{3}\langle P_2(\hat{H}\cdot\hat{e}'_s) - P_2(\hat{H}\cdot\hat{e}''_s)\rangle$, where $\hat{e}'_s, \hat{e}''_s$ denote the directions of the other two principal axes of that tensor. This contribution is usually omitted, being very small compared to the primary term in eq (11). It is also omitted in this work as it would only have insignificant quantitative effects on the final results.

The contribution of the Z-dependent part involves the biaxiality order parameter $\left\langle P_2(\hat{m}\cdot\hat{e}_s) - P_2(\hat{l}_h\cdot\hat{e}_s)\right\rangle$ which according to theoretical estimates[19] is rather small. In that case the spectral signature for perpendicular disposition of the helix axis is essentially one half of the signature for the parallel disposition. This is in accord with measured spectra obtained parallel and perpendicular to the magnetic field direction in the neat $N_X$ phase of CB9CB dimers.[13] Furthermore, rapid diffusion of the solute molecule along the Z direction within the NMR measurement time would lead to complete averaging out of the $\cos 2kZ$ factor in the measured spectra, thus resulting in an exact ½ shrinking of the perpendicular spectra relative to those measured for parallel alignment. We shall return to the consideration of the possible influence of diffusion on the measured spectra in section 6.

## 4. Solute symmetries and enantiotopic differentiation in the $N_X$ phase.

It has been shown[21] that in the conventional chiral nematic phase $N^*$ (i) the vector part $V^{(1)}$ of the potential of mean torque in eq (10) is absent by symmetry, (ii) the second rank part $V^{(2)}$ determines primarily the orientational ordering of the solute molecules but does not differentiate between enantiotopic sites and (iii) such differentiation results from the pseudovector contribution $V^{(p)}$ associated with the molecular chirality of the solvent (i.e. the counterpart of, $V^*(\omega, Z)$ in eq (10), adapted to the symmetries of the $N^*$ phase), and only for solutes belonging to the point group symmetries $C_s$, $C_{2v}$, $D_{2d}$ and $S_4$. Since the spectral signatures in eq(12) for the $N_X$ phase involve the same order parameters as in the case of the $N^*$ phase, it follows that the molecular chirality contribution $V^*(\omega, Z)$ in eq (10) would lead to enantiotopic discrimination in the $N_X$ phase as well for the same four point group symmetries of the solute molecules. The magnitude of such discrimination would depend on the extent of the induced chiral imbalance on the solvent molecules in the $N_X$ phase, unlike the $N^*$ phase wherein discrimination reflects directly the intrinsic molecular chirality. As the chiral imbalance is estimated[19] to be small in the $N_X$ phase, the resulting enantiotopic discrimination might be marginal. On the other hand, the vector term $V^{(1)}(\omega, Z)$ in eq (10) is present and estimated[19] to be strong in the $N_X$ phase and therefore could lead to appreciable enantiotopic discrimination albeit only for the point group symmetries which are compatible with the existence of polar asymmetry of the solute molecules. Thus, of the aforementioned four point group symmetries, only solutes of the $C_s$, $C_{2v}$ symmetries would show enantiotopic discrimination associated with the vector term $V^{(1)}(\omega, Z)$. These are considered separately below.

**4.1. Mirror plane ($C_s$).** With the z molecular axis taken as the plane normal, three of the five independent geometrical constants in eq (12) are identical for any pair of mirror-image sites $s, s'$ and the remaining two geometrical constants are of opposite sign, namely,

$$S_{xz}^{s'} = -S_{xz}^{s} \; ; \; S_{yz}^{s'} = -S_{yz}^{s} \quad . \tag{15}$$

The non-vanishing terms in the second rank part of the potential of mean torque, $V^{(2)}(\omega; Z)$ in eq (6), are those with the index pairs $ab = xx / yy / zz / xy$. For the polar part, $V^{(1)}(\omega; Z)$ in eq (4), the terms which could survive are those with $a = x, y$. Accordingly, the total potential of mean torque in eq (10), ignoring completely the molecular chirality contribution $V^*(\omega, Z)$, leads to non-vanishing values of the solute order parameters $\langle S_{xz}^{n_h} \rangle$ and $\langle S_{yz}^{n_h} \rangle$ appearing in eq (12) and therefore to the spectral difference between the $s, s'$ enantiotopic molecular elements. In quantitative terms:

$$(\nu_s - \nu_{s'})_\parallel / \nu_s^0 = \frac{8}{3}\left(S_{xz}^{s}\langle S_{xz}^{n_h}\rangle + S_{yz}^{s}\langle S_{yz}^{n_h}\rangle\right) \tag{16}$$

**4.2. Two mirror planes containing a twofold symmetry axis ($C_{2v}$).** With the z molecular axis taken as the $C_2$ axis, and the x, y axes on the two mirror planes, we have for a pair of sites $s, s'$ which are enantiotopic with respect to that plane,

$$S_{xy}^{s'} = -S_{xy}^{s} \tag{17}$$

As the solute order parameters $\langle S_{xz}^{n_h}\rangle, \langle S_{yz}^{n_h}\rangle$ vanish by symmetry, the geometrical constants $S_{xz}^{s}, S_{yz}^{s}$ are of no relevance to the expression for the spectral signatures in eq(12) and the remaining two geometrical constants in that expression are identical for the two sites. The non-vanishing terms in the 2$^{nd}$ rank part of the potential of mean torque, are those with the index pairs $ab = xx / yy / zz$. The surviving term for the polar part is the one with $a = z$ which, together with the non-vanishing of the $\langle \Phi_{xx}^{(2)} \rangle', \langle \Phi_{yy}^{(2)} \rangle'$ contributions imply the non-vanishing of the solute order parameter $\langle S_{xy}^{n_h} \rangle$. Therefore the spectral differentiation between enantiotopic sites is given by

$$\left(\nu_s - \nu_{s'}\right)_{\parallel} / \nu_s^0 = \frac{8}{3} S_{xy}^{e_s} \langle S_{xy}^{n_h} \rangle \quad . \tag{18}$$

In summary, ignoring any chiral asymmetry of the solvent molecules, enantiotopic discrimination can be observed on the NMR spectra of small rigid solutes in $N_X$ solvents only for solutes of molecular symmetry corresponding to one of the two point groups, $C_s$ or $C_{2v}$.

## 5. Inferences for the potential of mean torque from NMR measurements on small rigid solutes in the N and $N_X$ phases

A limited number of experimental NMR studies on small rigid solutes in $N_X$ solvents are presently available in the literature.[12,22–24] These include primarily the polar and prochiral solute acenaphthene[12], which has also been subject to extensive experimental NMR studies in the $N^*$ phase[25] and found to exhibit clear enantiotopic spectral discrimination in both the $N_X$ and the $N^*$ phases. As the present work is mainly focused on enantiotopic discrimination, which is a distinguishing feature of the $N_X$ phase with respect to the higher temperature N phase, most of our attention will be devoted to acenaphthene. However, measurements are also available for more symmetric rigid solutes in the $N_X$ phase, namely p-dichloro- and trichloro-benzene[22] and anthracene.[23] Although these molecules, being apolar and containing no enantiotopic elements, offer little information on the distinguishing features of the $N_X$ phase, they will be briefly considered here to demonstrate the consistent transferability of the proposed form of the potential of mean torque. Experimental studies on flexible solutes in the $N_X$ phase are also available[12,26] but our consideration is restricted here to rigid solutes in order to avoid complexities originating from the flexibility. Flexible solutes in the $N_X$ phase will be addressed in a forthcoming communication.

### 5.1. Polar prochiral solutes: Acenaphthene.

Acenaphthene has $C_{2v}$ point group symmetry (see Figure 2a); accordingly the expression for the spectral signatures of eq (12) reduces to

$$\left(\nu_s / \nu_s^0\right)_{\parallel} = S_{zz}^s \langle S_{zz}^{n_h} \rangle + \frac{1}{3}\left(S_{xx}^s - S_{yy}^s\right)\left(\langle S_{xx}^{n_h} \rangle - \langle S_{yy}^{n_h} \rangle\right) + \frac{4}{3} S_{xy}^s \langle S_{xy}^{n_h} \rangle \tag{19}$$

The deuterium quadrupolar spectra of acenaphthene as a solute in the $N_X$ phase have been studied experimentally[12] and found to show enantiotopic discrimination for the prochiral pairs 19, 20 and 20, 21 in the numbering of Figure 2a. As we are dealing with quadrupolar splittings, $\Delta\nu_{CD}^s$, corresponding to the deuterated sites $s=13$ to $22$, from magnetically aligned samples, the relevant spectral signatures of eq (19) are related to the measurable splittings as $\left(v_s/v_s^0\right)_{\parallel} = \left(\frac{3}{2}q_{CD}^s\right)^{-1}\Delta v_{CD}^s$, where the qudrupolar coupling constant has the value $q_{CD}^s \approx 167 kHz$ for aliphatic C-D bonds (in the present case for sites $s=19$ to $22$) and $q_{CD}^s \approx 190 kHz$ for aromatic ones (sites $s=13$ to $18$). Due to the twofold rotational symmetry about the $z$-axis there are only three distinct splittings for the aromatic sites ("in-plane" sites) $\Delta v_{CD}^{13} = \Delta v_{CD}^{17}$, $\Delta v_{CD}^{14} = \Delta v_{CD}^{18}$, $\Delta v_{CD}^{15} = \Delta v_{CD}^{16}$, and two distinct splittings for the out-of-plane (enantiotopic, aliphatic) sites $\Delta v_{CD}^{19} = \Delta v_{CD}^{21}$ and $\Delta v_{CD}^{20} = \Delta v_{CD}^{22}$, in the $N_X$ phase. The latter two splittings collapse into a single splitting in the N phase where the phenomenon of enantiotopic discrimination is not observed experimentally.

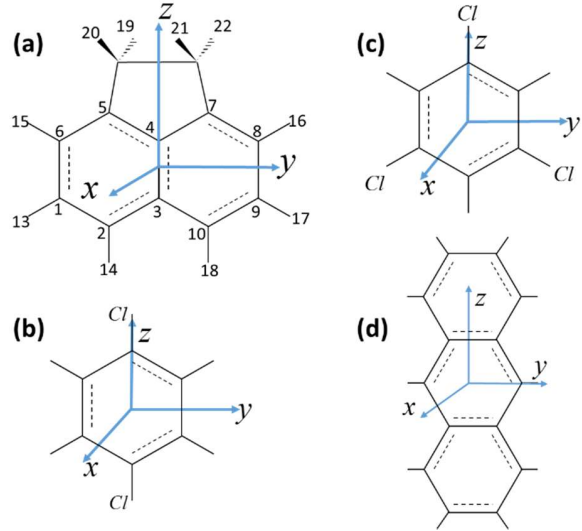

**Figure 2**: Molecular diagrams showing the assignment of axes for (a) acenaphthene, showing also the site labeling used in the text, (b) *para*-dichlorobenzene (*p*-dcb), (c) trichlorobenzene (tcb) and (d) anthracene. Note that the choice of axes is such that the $z$ molecular axis is in all cases an axis of twofold symmetry and the $x$ axis is normal to the molecular plane.

It follows directly from eq (19) and the symmetry considerations in section 4.2, that the temperature dependence of the three splittings associated with the in-plane sites and of the average of the splittings associated with the out-of-plane sites $\overline{\Delta v_{CD}^{19,20}} = \left(\Delta v_{CD}^{19} + \Delta v_{CD}^{20}\right)/2$, is determined by just two independent order parameters of the solute molecule, namely $\langle S_{zz}^{n_h} \rangle$ and $\left(\langle S_{xx}^{n_h} \rangle - \langle S_{yy}^{n_h} \rangle\right)$. Accordingly, two of these four splittings, say those of $s=14$ and $15$, suffice for the determination of the two order parameters and the remaining two splittings, $\Delta v_{CD}^{13}$ and $\overline{\Delta v_{CD}^{19,20}}$, will be expressible as linear combinations of the first two, with constant coefficients. In this case we obtain the following relations between the order parameter and the splittings:

$$\left\langle S_{zz}^{n_h} \right\rangle = \frac{\Delta v_{CD}^{15}(1-S_{zz}^{14}) - \Delta v_{CD}^{14}(1-S_{zz}^{15})}{\frac{3}{2} q_{CD}^{aromatic} \left( S_{zz}^{15} - S_{zz}^{14} \right)}; \quad \left( \left\langle S_{xx}^{n_h} \right\rangle - \left\langle S_{yy}^{n_h} \right\rangle \right) = 2 \frac{\Delta v_{CD}^{15} S_{zz}^{14} - \Delta v_{CD}^{14} S_{zz}^{15}}{q_{CD}^{aromatic} \left( S_{zz}^{15} - S_{zz}^{14} \right)} \quad (20)$$

and the following two linear relations among splittings:

$$\left( \frac{\Delta v_{CD}^{13}}{\Delta v_{CD}^{14}} \right) = \left( \frac{S_{zz}^{13} - S_{zz}^{14}}{S_{zz}^{15} - S_{zz}^{14}} \right) + \left( \frac{\Delta v_{CD}^{15}}{\Delta v_{CD}^{14}} \right) \left( \frac{S_{zz}^{15} - S_{zz}^{13}}{S_{zz}^{15} - S_{zz}^{14}} \right) \quad (21)$$

and

$$\left( \frac{\overline{\Delta v_{CD}^{19,20}}}{\Delta v_{CD}^{14}} \right) \left( \frac{q_{CD}^{aromatic}}{q_{CD}^{aliphatic}} \right) = \left( \frac{S_{zz}^{19,20}(1-S_{zz}^{15}) + (S_{xx}^{19,20} - S_{yy}^{19,20})S_{zz}^{15}}{S_{zz}^{15} - S_{zz}^{14}} \right) - \left( \frac{\Delta v_{CD}^{15}}{\Delta v_{CD}^{14}} \right) \left( \frac{S_{zz}^{19,20}(1-S_{zz}^{14}) + (S_{xx}^{19,20} - S_{yy}^{19,20})S_{zz}^{14}}{S_{zz}^{15} - S_{zz}^{14}} \right) \quad (22)$$

It should be noted that the relations in eqs(21) to (22) are expected to hold in the N, $N_X$ and $N^*$ phases, assuming that the geometry of the rigid acenaphthene solute remains unchanged in the different solvent phases.

The temperature dependence of the order parameters in eq(20), as obtained from the measured splittings of acenaphthene in the N and $N_X$ phases, and using the geometrical identity $\left\langle S_{xx}^H \right\rangle + \left\langle S_{yy}^H \right\rangle + \left\langle S_{zz}^H \right\rangle = 0$, is shown in Figure 3. It is evident from these graphs that the most ordered molecular axis is the *x* axis, i.e. the axis normal to the planar core of the acenaphthene molecule and is ordered at right angles to the direction of $\hat{H}$ (negative values of the order parameter $\left\langle S_{xx}^H \right\rangle$) while the least ordered axis is the twofold symmetry axis *z*. All three order parameters show very limited variation with temperature in the $N_X$ phase.

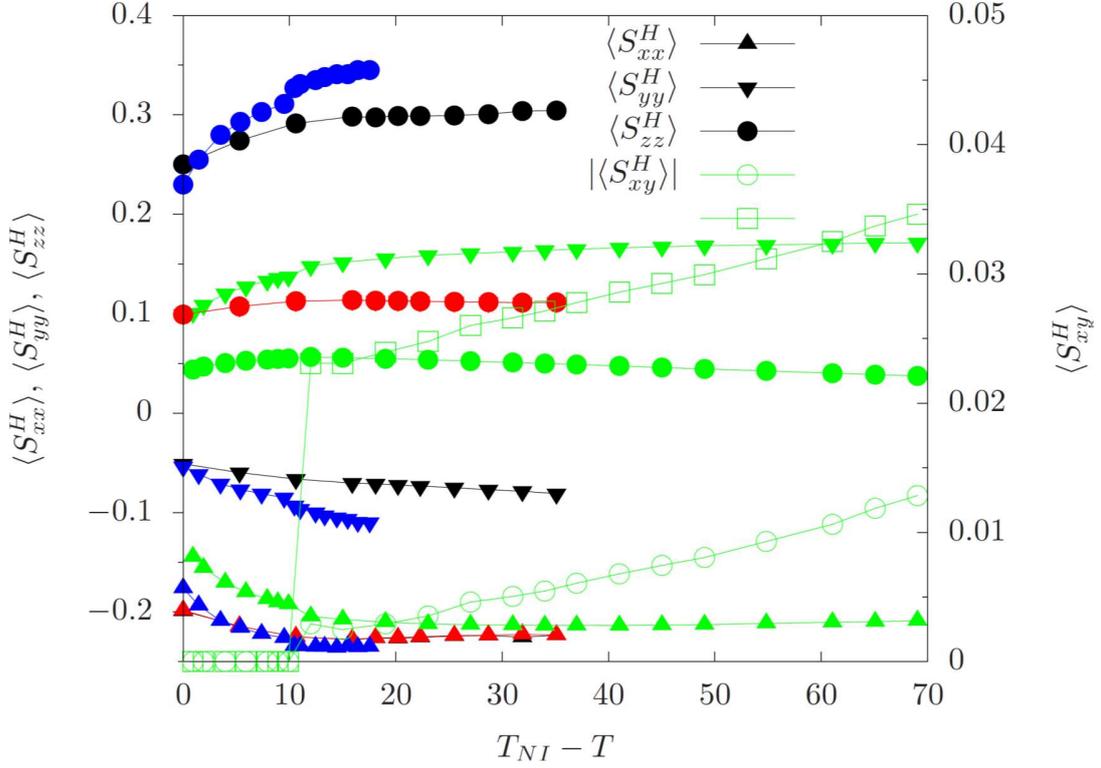

**Figure 3.** Calculated values of the order parameters $\langle S_{xx}^H \rangle, \langle S_{yy}^H \rangle, \langle S_{zz}^H \rangle$ and $\langle S_{xy}^H \rangle$ for all the rigid solutes studied in the in the N and N$_X$ phases. The direction of the magnetic field $\hat{H}$ coincides with the direction of the helical axis $\hat{n}_h$ in the case of the N$_X$ phase and with the nematic director $\hat{n}$ in the case of the N phase. The molecular axes $x, y, z$ for the various solutes are assigned as shown in Figure 2. The color code used in the graphs is: acenaphthene in dimer CB7CB (green), p-dichlorobenzene (black) and trichlorobenzene (red) in CBC9CB/5C mixtures and anthracene (blue) in CB7CB. The calculated values of the $\langle S_{xx}^H \rangle$ order parameter for p-dichlorobenzene and trichlorobenzene practically coincide. Note the different scale on the right hand side vertical axis for the order parameter $\langle S_{xy}^H \rangle$, applicable only to the acenaphthene solute. The two plots for this order parameter are obtained according to eq (23) (open circles), based on perfectly rigid molecular geometry and eq (26) (open squares), which allows for small deformations of the aliphatic part.

The order parameter $\langle S_{xy}^H \rangle$ has non-vanishing values only in the N$_X$ phase; they are obtained according to eq(19) from the experimental splittings of the enantiotopic sites as:

$$\langle S_{xy}^H \rangle = \left( \Delta \nu_{CD}^{19} - \Delta \nu_{CD}^{20} \right) / 4 S_{xy}^{19} q_{CD}^{aliphatic} \qquad (23)$$

The temperature dependence of the so obtained $\langle S_{xy}^H \rangle$ is plotted in Figure 3. The absolute values of this, as well as of the other three acenaphthene order parameters shown in Figure 3 are larger than the respective order parameters of the same solute in the N* phase of the PBLG/CHCl$_3$ solvent[21] by two orders of magnitude; naturally, this is also the case for the respective splittings. To obtain some insight into the mechanisms giving rise to the enantiotopic discrimination in the two solvents, we have compared the relative strength of the effect, i.e. the frequency ratios

$$\rho_{CD}^{19,20} = \left| \left( \Delta \nu_{CD}^{19} - \Delta \nu_{CD}^{20} \right) / \left( \Delta \nu_{CD}^{19} + \Delta \nu_{CD}^{20} \right) \right| , \qquad (24)$$

of the acenaphthene solute in the N* and N$_X$ solvents. The temperature dependence of these ratios is shown in Figure 4. It is apparent from these plots that, while the relative intensities of the enantiotopic discrimination are of the same order of magnitude, the qualitative trends of the temperature dependence the frequency ratios differ substantially.

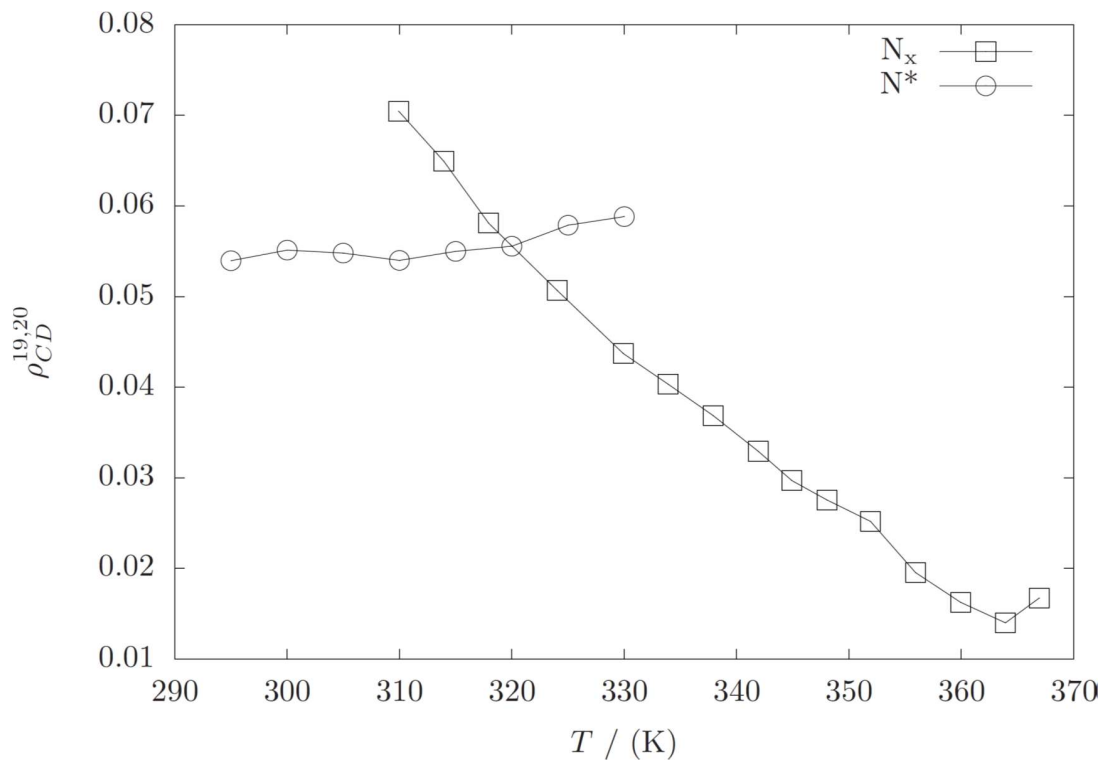

**Figure 4.** Temperature dependence of the relative strength of spectral enantiotopic differentiation, quantified by means of the splitting ratio in eq (24), for acenaphthene in the N$_X$ phase of the mesogenic dimer CB7CB[12] and in the N* phase of PBLG/CHCl$_3$[25].

The ratio plots of the splittings corresponding to eqs (21) and (22) are shown in Figure 5 for the acenaphthene solute in N, N$_X$ and N* solvents. Evidently, the experimental splittings in Figure 5(a) satisfy rather accurately the linear dependence deriving from eq (21) throughout the temperature range of the N, $N_X$ and N* phases. The respective slope is 0.938 and the intercept 0.069. These practically coincide with the theoretical values (slope =0.934; intercept=0.067) obtained from the geometrical parameters[27] $S_{zz}^{13} = -0.199, S_{zz}^{14} = 0.994, S_{zz}^{15} = -0.120$ in eq(21).

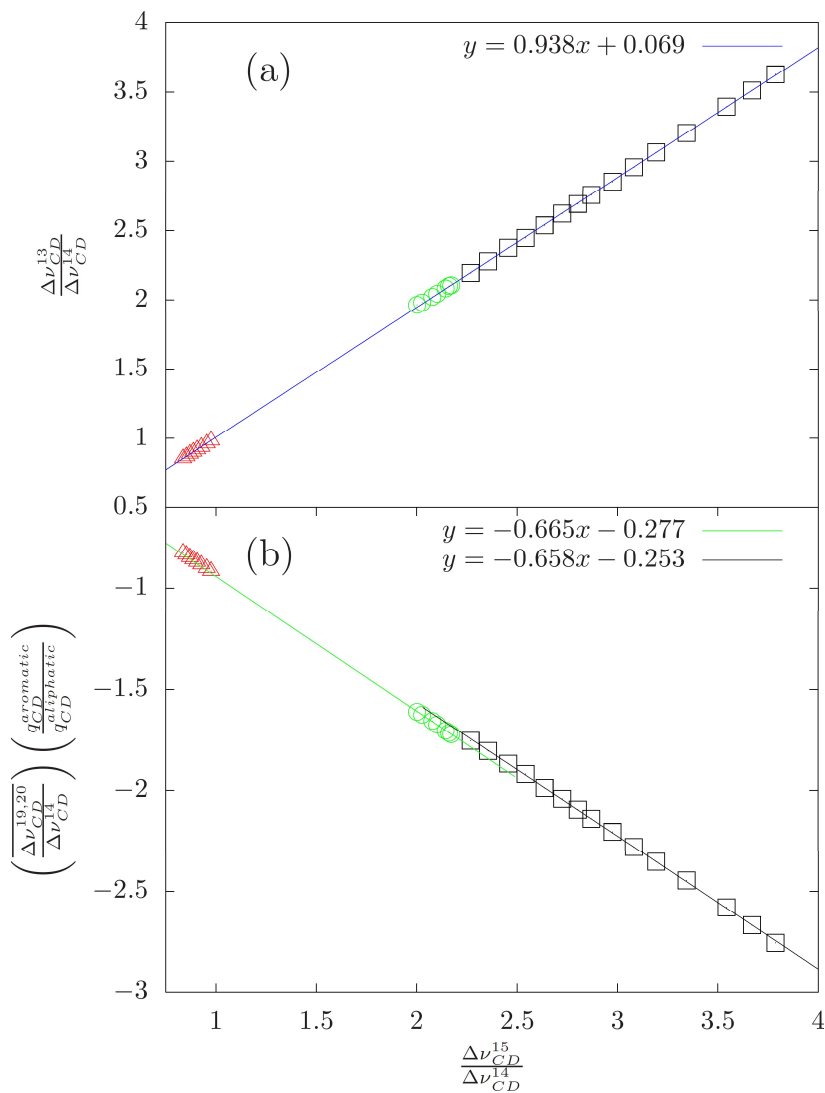

**Figure 5.** Ratio plots of measured splittings of acenaphthene dissolved in the N (circles) and $N_X$ (squares) phases of dimer CB7CB[12] and the N* (triangles) phase of PBLG/CHCl$_3$[25]. The upper graph (a) corresponds to the in-plane sites appearing in the linear relation of eq (21) and lower graph (b) to the out-of-plane and in-plane sites in eq (22).

The linear dependence is also satisfied by the experimental splittings in Figure 5(b); however there is a small but clear change in the slope and the intercept of the linear plots across the N to $N_X$ phase transition. Thus the experimental slope in the N phase range of Figure 5(b) is $-0.665$ and the intercept is $-0.277$, the respective values are $-0.658$ and $-0.253$ in the $N_X$ phase and the theoretical values, obtained from eq (22) with $S_{zz}^{19,20} = -0.157$, $(S_{xx}^{19,20} - S_{yy}^{19,20}) = 0.775$ [27] are $-0.646$ and $-0.287$, respectively.

A possible interpretation of this observation is based on the hypothesis that the chiral environment experienced by the acenaphthene solute in an $N_X$ domain induces small deformations on the aliphatic part of the molecule, while leaving the relatively stiffer planar aromatic core essentially unaffected. Assuming, for example, that the deformation consists of a mere rotation of the C11-C12 bond direction (see Figure 2a) about the z molecular axis by a small angle $\alpha$, whose sign is inverted on inverting the twisting sense of the domain, this deformation would modify eqs (22) and (23) into

$$\left[\left(\frac{\overline{\Delta v_{CD}^{19,20}}}{\Delta v_{CD}^{15}}\right)\left(\frac{q_{CD}^{aromatic}}{q_{CD}^{aliphatic}}\right) - S_{zz}^{19}\frac{(1-S_{zz}^{14}) - (\Delta v_{CD}^{14}/\Delta v_{CD}^{15})(1-S_{zz}^{15})}{(S_{zz}^{15} - S_{zz}^{14})}\right]\frac{\cos 2\alpha}{S_{xx}^{19,20} - S_{yy}^{19,20}} =$$
$$= S_{zz}^{14} - (\Delta v_{CD}^{14}/\Delta v_{CD}^{15})S_{zz}^{15} + \left(\frac{q_{CD}^{aromatic}}{4S_{xy}^{19}q_{CD}^{aliphatic}}(\Delta v_{CD}^{19} - \Delta v_{CD}^{20})/\Delta v_{CD}^{15}\right)\sin 2\alpha$$
(25)

and

$$\langle S_{yz}^n \rangle = \frac{(\Delta v_{CD}^{19} - \Delta v_{CD}^{20})}{4 q_{CD}^{aliphatic} S_{xy}^{19} \cos 2\alpha} + \frac{1}{2}\left(\langle S_{xx}^n \rangle - \langle S_{yy}^n \rangle\right)\tan 2\alpha \quad . \tag{26}$$

Optimizing the agreement of eq (25) with the measured splittings in the $N_X$ phase yields an essentially constant value of the deformation angle $\alpha \approx 3.5°$, for which the experimental points are reproduced by eq (25) to within experimental accuracy. The respective values of the order parameter $\langle S_{yz}^{n_h} \rangle$, calculated according to eq (26), are shown in the plots of Figure 3, where it is evident that they differ appreciably from the values calculated according to the "deformation-free" eq (23).

Next we consider the implications of the above NMR observations for the potential of mean torque of eq (10) in the case of the acenaphthene solute in the N and $N_X$ phases. According to the $C_{2v}$ symmetry of the acenaphthene molecules and the choice of molecular axes shown in Figure 2a, the polar contribution in eq (4) is

$$V^{(1)}(\omega; Z) = u_{polar}(\hat{z} \cdot \hat{m}) \quad . \tag{27}$$

Dropping, for notational simplicity, the subscripts in $n_h$ and $l_h$ when describing the parameters of the potential, with the understanding that the helix axis of the $N_X$ phase is replaced by the director $\hat{n}$ in the uniaxial N phase (wherein the $\hat{l}_h$ becomes irrelevant), the explicit expression for $V^{(2)}(\omega;Z)$ of eq (6) consists of the terms:

$$V^{(2)}(\omega;Z) = u_0^n S_{zz}^n + \frac{2}{3}u_2^n\left(S_{xx}^n - S_{yy}^n\right) +$$
$$+ \frac{2}{3}b_0^{lm}\left(S_{zz}^l - S_{zz}^m\right) + \frac{2}{3}b_2^{lm}\left(S_{xx}^l - S_{xx}^m - S_{yy}^l + S_{yy}^m\right) + \quad . \tag{28}$$
$$+ 2\varepsilon_0^{nl}q_{zz}^{nl} + \varepsilon_2^{nl}\left(q_{xx}^{nl} - q_{yy}^{nl}\right)$$

In the above equations the parameter $u_{polar}$ describes the polar orientational coupling of the molecular $z$ axis to the local polar director $\hat{m}$ of the $N_X$ phase, $u_0^n$ describes the apolar orientational coupling of the same axis to the helix direction $\hat{n}_h$, $u_2^n$ describes the difference in such coupling between the $x$ and $y$ axes (molecular biaxiality). The parameters $b_0^{lm}, b_2^{lm}$ describe the coupling of the $z$ axis and $x$-$y$ difference, respectively, to the local biaxiality of the phase about the helix direction $\hat{n}_h$ and $\varepsilon_0^{lm}, \varepsilon_2^{lm}$ describe the deviation (described as eccentricity or tilt in sections 1 and 2) of the maximal ordering of the axes $z$, $x$-$y$, respectively, from the direction $\hat{n}_h$. The explicit relation of all the above parameters to the solvent order parameters and the molecular coupling parameters of eq (8) is given in eq (AII.2') of Appendix II. It is noted here that the signs of both $\varepsilon_0^{lm}$ and $\varepsilon_2^{lm}$ are inverted[19] on inverting the sense of twisting of the $\hat{m}$ director, whereas the other four parameters in eq (28) and $u_{polar}$ of eq (27) have identical values for domains of either twisting sense. As shown below, it is the combination $u_{polar}\varepsilon_2^{lm}$ that gives rise to the enantiotopic discrimination in the $N_X$ phase. This combination of the polar ordering along the director $\hat{m}$ with the eccentric/tilted ordering induced on the non-polar molecular axes $x, y$ of the solute in the $\hat{n}_h - \hat{l}_h$ plane will be referred to as "structural chirality" of the solvent, to emphasize its distinction from (i) "molecular chirality", which could be present in the $N_X$ solvent as a result of a possible chiral asymmetry induced on the solvent molecules and (ii) "modulation chirality" associated with the tight twisting of the $\hat{m}$ director about $\hat{n}_h$.

The part of the potential of mean torque associated with the presence molecular chirality in the $N_X$ solvent, eq (9), for a solute molecule of $C_{2v}$ symmetry, using a right handed molecular frame as shown in Figure 2a for acenaphthene, has the following terms:

$$\begin{aligned}V^*(\omega;Z) = &-2c^{*n}\left((y\cdot\hat{n}_h)(x\cdot\hat{n}_h)\right)\\&-2c^{*lm}\left((y\cdot\hat{l}_h)(x\cdot\hat{l}_h)-(y\cdot\hat{m})(x\cdot\hat{m})\right)\\&+2c_+^{*nl}\left((x\cdot\hat{n}_h)(y\cdot\hat{l}_h)+(x\cdot\hat{l}_h)(y\cdot\hat{n}_h)\right)\\&+c_-^{*nl}\left((x\cdot\hat{n}_h)(y\cdot\hat{l}_h)-(x\cdot\hat{l}_h)(y\cdot\hat{n}_h)\right)\end{aligned} \qquad (29)$$

All of the above terms describe the product-ordering of the two nonpolar axes $(x,y)$ along the three local axes of the phase. The first term refers to the ordering along $\hat{n}_h$ and gives rise directly to a nonzero value of the solute order parameter $\langle S_{xy}^{n_h}\rangle$, thereby producing enantiotopic discrimination. The second term describes the biaxiality of the product ordering about $\hat{n}_h$ and the third and fourth terms represent the symmetric and antisymmetric combinations of the tilted ordering of the two molecular axes in the $\hat{n}_h - \hat{l}_h$ plane. The explicit relation of these parameters to the solvent chiral order parameters and molecular coupling parameters is given in eq (AII.3') of Appendix II.

Obviously, the three independent order parameters (Figure 3) that can be extracted directly from the NMR measurements are not sufficient for the complete determination of all the terms in the potential of mean torque, despite the restriction to only leading rank contributions for the latter. On the other hand, not all of these terms have a direct influence on the spectral signatures in eq (19), as the three molecular order parameters that determine these signatures are invariant with respect to rotations about the macroscopic $Z$ axis (in this case coinciding with the helical axis $\hat{n}_h$ about which the $\hat{m}$ director twists). Accordingly, those terms of the potential of mean torque which do depend on rotations about the $Z$ axis will have an indirect influence on the spectrum, through the averaging of the corresponding orientational distribution with respect to rotations about $Z$. To illustrate this, we describe the orientation of the solute molecular frame $x,y,z$ relative to the macroscopic frame $X,Y,Z$ by the Euler angles $\varphi,\theta,\psi$. Then, the complete potential of mean torque $V(\omega,Z)$ of eq (10) can be expressed in terms of $\theta,\psi$ and a third angle $\tilde{\varphi}$ conveying the combination of the twisting angle $k(Z-Z_0)$ of eq (1) with the Euler angle $\varphi$, as $\tilde{\varphi} = \varphi - k(Z-Z_0)$; that is $V(\omega,Z) = V(\theta,\psi;\tilde{\varphi})$, wherein the $Z$-dependence is now absorbed in the angle $\tilde{\varphi}$. Accordingly, the ensemble average $\langle Q \rangle$ of a quantity $Q(\theta,\psi)$, such as any of the three order parameters in eq (19), would be obtained in terms of a "rotationally averaged" potential of mean torque $\bar{V}(\theta,\psi)$ as $\langle Q \rangle = (1/\zeta)\int \sin\theta d\theta d\psi Q(\theta,\psi)e^{\bar{V}(\theta,\psi)}$. As shown in Appendix II, the leading rank contributions of $\bar{V}(\theta,\psi)$ for the acenaphthene solute in the $N_X$ phase are just three:

$$\bar{V}(\theta,\psi) \approx u_0 \left(\frac{3}{2}\cos^2\theta - \frac{1}{2}\right) + u_2 \sin^2\theta \cos 2\psi + u_1 \sin^2\theta \sin 2\psi \quad , \tag{30}$$

where the parameters $u_0, u_2, u_1$ combine direct contributions from the $\tilde{\varphi}$-independent terms of $V(\theta,\psi;\tilde{\varphi})$ as well as indirect contributions from the $\tilde{\varphi}$-dependent part (see eq (AII.6). The latter contributions are quadratic in the coupling parameters that describe the $\tilde{\varphi}$-dependent part. The values of these parameters are estimated to be well below 1, rendering their quadratic contributions substantially smaller than their direct contributions in eqs (27), (28) and (29).

In the uniaxial apolar nematic phase N, $u_1$ vanishes by symmetry and the solute ordering is described completely by the $u_0, u_2$ terms which reduce to $u_0 = u_0^n$, $u_2 = u_2^n$. In the $N_X$ phase, it follows from eqs (AII.6) of Appendix II that, if the molecular chirality of the solvent is ignored completely and the quadratic contributions are neglected compared to the linear contributions $u_0^n$ and $u_2^n$, then $u_0 \approx u_0^n$, $u_2 \approx u_2^n$ and $u_1 \approx \frac{1}{2}\varepsilon_2^{nl} u_{polar}$ i.e. the parameters $u_0, u_2$ are dominated by their N phase expressions while $u_1$ is produced by the simultaneous presence of (i) the polar ordering of the solute $z$ axis along the polar director $\hat{m}$ and (ii) the tilted ordering of the non-polar molecular axes $x, y$ in the plane of the (also non-polar) $\hat{n}_h - \hat{l}_h$ phase axes. In other words, the $u_1$ contribution, which produces the non-vanishing values of the order parameter $\langle S_{xy}^{n_h} \rangle$, and thereby gives rise to the enantiotopic spectral discrimination, is generated by the "structural chirality" of the $N_X$ phase. Figure 6 shows the temperature dependence of the optimal values of the parameters $u_0, u_2, u_1$ which are obtained from fits to the experimentally determined acenaphthene solute order parameters in Figure 3.

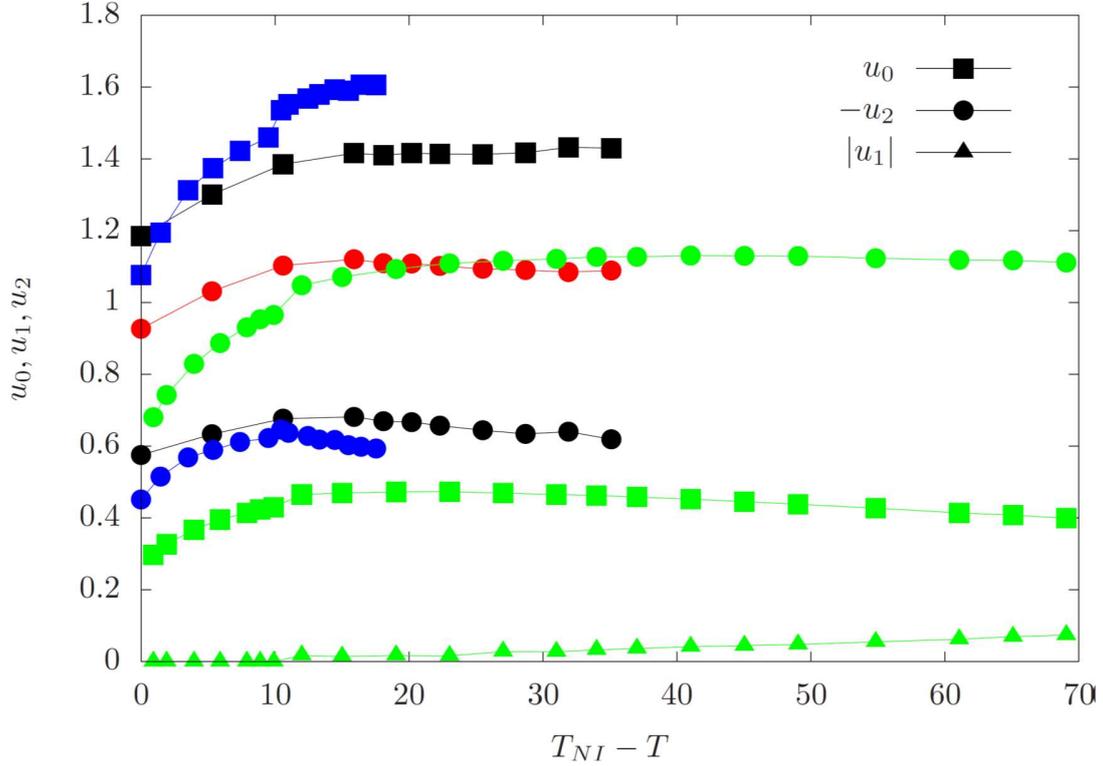

**Figure 6.** Temperature dependence of optimal values of the parameters $u_0, u_2, u_1$ of the potential of mean torque i) of eq (30) for acenaphthene (green); ii) of eq (34) for *p*-dcb (black) and for anthracene (blue) and iii) of eq (37) for tcb (red). The optimal values are obtained by fitting the respective order parameters in Figure 3.

In accord with the $x$ axis (molecular plane normal) being the most ordered of the three molecular axes, the parameter $u_2$ shows the largest absolute values and is negative. Both $u_0$ and $u_2$ show an abrupt change across the N to $N_X$ phase transition and on further reducing the temperature the $u_0$ values show a gradual decrease while those of $u_2$ remain practically constant. The decrease of $u_0$ can be rationalized by considering its composition in eq (AII.6) and noting that the increasing tendency of the polar $z$ molecular axis to align with the polar director $\hat{m}$, rather than the helix direction $\hat{n}_h$, implies a weakening of the linear contribution $u_0^n$. In addition, a number of quadratic terms, notably $(\varepsilon_2^{nl})^2$ and $(u_{polar})^2$, enter with a negative sign in eq (AII.6); the growth of such terms on lowering the temperature could contribute further to the overall decrease of the composite parameter $u_0$. The $u_1$ parameter acquires non vanishing values below the N to $N_X$ transition temperature and shows a gradual increase with decreasing temperature, indicating essentially the increase of the product $\varepsilon_2^{nl} u_{polar}$. However, $u_1$ remains smaller than the absolute values of $u_0$ and $u_2$ by at least an order of magnitude at all temperatures.

In all of the above analysis, the size of the acenaphthene solute molecules is neglected relative to the pitch of the helical twisting in the solvent phase. According to the formulation in appendix I, the spatial extension of a rigid solute introduces corrections on the order of $\sin(k(\hat{n}_h \cdot \vec{r}))$ etc. Now, for typical N$_X$ phases, the pitch is on the order of $2\pi/k \sim 10\,\text{nm}$ while the partitioning of the acenaphthene molecule into two mirror-image halves on either side of the *x-z* plane (see Figure A1) introduces an intramolecular distance of the order of $r \sim 10^{-1}\,\text{nm}$. Therefore the relative magnitude of the corrections would be around 5-10%, which is marginal, particularly in view of the further rotational averaging into which the $\tilde{\varphi}$-dependent part of the potential of mean torque is subjected in order to yield the effective parameterization of eq (30).

**5.2. Apolar planar solutes.**

To our knowledge, the experimental NMR studies of small rigid solutes in the N$_X$ phase that are presently available in the literature include, aside from acenaphthene, apolar planar solutes, with no prochiral elements, and are exhausted with the cases of (i) dichloro- and trichloro-benzene, for which dipolar couplings in the N and N$_X$ phases of CBC9CB/5C mixtures have recently been studied by NMR[22] and have been analyzed using a different approach from the one we present here and (ii) anthracene, for which deuterium quadrupolar splittings were studied in the N and N$_X$ phases of CB7CB[23] long before the conclusive identification of the N$_X$ phase, at which time the phase appearing on lowering the temperature from the conventional nematic was reported in that study to be a smectic.

Below, we describe these spectra in terms of the respective forms of the potential of mean torque derived from the one used for acenaphthene (eqs (27) and (28)) by imposing the additional symmetries of these planar apolar rigid solutes.

**5.2.1. The *para*-dichlorobenzene (*p*-dcb)** molecule has 3 mirror planes and 3 twofold axes normal to them. Choosing the molecular frame so that the *x*-axis is normal to the ring plane and the *z*-axis is along the Cl-Cl molecular direction (Figure 2b), we have the independent inversion symmetries $x \leftrightarrow -x,\ y \leftrightarrow -y,\ z \leftrightarrow -z$. Accordingly, of the molecular order parameters appearing in eq (12), only $\langle S_{zz}^H \rangle$ and $\langle S_{xx}^H \rangle - \langle S_{yy}^H \rangle$ survive. The molecule has three inequivalent types of proton pairs. One corresponds to adjacent (*ortho*) protons, with its inter-nuclear vector along the *z*-axis, another type is that of second neighbor (*meta*) protons, with the inter-nuclear vector along the *y*-axis and the third type is that of diametrically opposite (*para*) protons, making $\pi/3$, $\pi/6$ angles with the *z, y* axes respectively. Thus the relevant molecular geometry parameters which correspond to the various pairs have the values:

$$S_{xx}^{ortho} = S_{xx}^{meta} = S_{xx}^{para} = -1/2$$
$$S_{zz}^{ortho} = 1;\ S_{zz}^{meta} = -1/2;\ S_{zz}^{para} = -1/8$$
$$\left(S_{xx}^{ortho} - S_{yy}^{ortho}\right) = 0;\ \left(S_{xx}^{meta} - S_{yy}^{meta}\right) = -3/2 \qquad (31)$$
$$\left(S_{xx}^{para} - S_{yy}^{para}\right) = -9/8$$

and we get from eq (12) the following expressions for the spectral signatures of the proton pairs in terms of the two order parameters of the solute molecule:

$$\begin{aligned} \nu_s / \nu_s^0 &= \langle S_{zz}^H \rangle && \text{for } s = ortho \\ &= -\frac{1}{2}\langle S_{zz}^H \rangle - \frac{1}{2}\left(\langle S_{xx}^H \rangle - \langle S_{yy}^H \rangle\right) && \text{for } s = meta \\ &= -\frac{1}{8}\langle S_{zz}^H \rangle - \frac{3}{8}\left(\langle S_{xx}^H \rangle - \langle S_{yy}^H \rangle\right) && \text{for } s = para \end{aligned} \qquad (32)$$

Eliminating the molecular order parameters from these equations, we obtain a linear relation among the spectral signatures of the three pair types. Recalling that, in the case of dipolar couplings $D_{ij}$, the spectral signatures are given by $\left(\nu/\nu^0\right)_{ij} = -D_{ij} r_{ij}^3 / K_{p-p}$,[21] this linear relation is written in terms of the experimentally measured dipolar couplings $D_{ij}$ of the three pairs as follows:

$$D_{meta} / D_{para} = \frac{4}{3}\left(r_{para}/r_{meta}\right)^3 - \frac{1}{3}\left(r_{ortho}/r_{meta}\right)^3 D_{ortho}/D_{para} \quad . \qquad (33)$$

The experimental ratio plot in Figure 7 shows that the linear relation of eq (33) holds quite accurately at all temperatures. The slope ($-0.064$) and the intercept ($1.88$) of the linear plot are fairly close to the theoretical values found from semi-empirical quantum chemistry calculations[27] of the intramolecular proton-proton distance ratios, namely $\frac{4}{3}\left(r_{para}/r_{meta}\right)^3 = 2.026$ and $\frac{1}{3}\left(r_{ortho}/r_{meta}\right)^3 = 0.061$. However, the two sets of values are not coincident and the order parameter calculations were done using the distance ratios obtained from the experimental plot of Figure 7. Figure 3 shows the temperature dependence of the molecular order parameters $\langle S_{xx}^H \rangle, \langle S_{yy}^H \rangle, \langle S_{zz}^H \rangle$ obtained from the experimentally measured spectral values according to eq (32). Notably, these order parameters show a continuous variation with temperature across the N to $N_X$ phase transition.

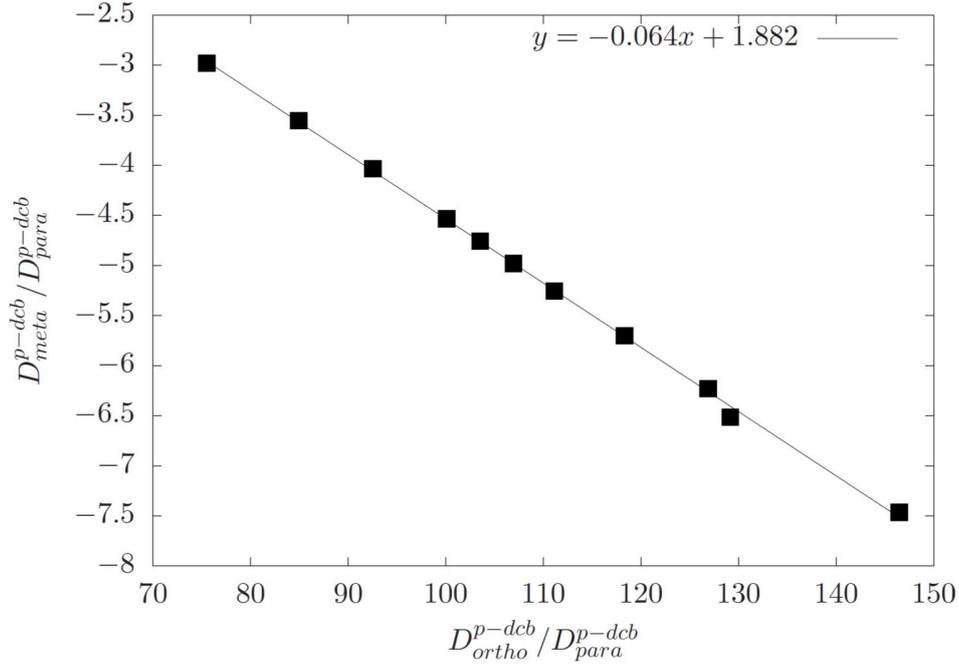

**Figure 7**. Plot of the ratios of measured dipolar coupling of *p*-dcb in the N and $N_X$ phases of a CBC9CB/5C mixture showing a linear relation, in accord with eq (33).

Due to the symmetry of the *p*-dcb molecule, its potential of mean torque as a solute in the $N_X$ phase has no polar or chiral parts; it therefore has $u_1 = 0$ and consists of the first two contributions in eq (30), i.e.

$$\bar{V}(\theta,\psi) \approx u_0 \left( \frac{3}{2}\cos^2\theta - \frac{1}{2} \right) + u_2 \sin^2\theta \cos 2\psi \quad , \tag{34}$$

with $u_0 = u_0^n + \frac{2}{21}(\varepsilon_0^{nl})^2 - \frac{1}{14}(\varepsilon_2^{nl})^2 - \frac{4}{21}(b_0^{lm})^2 + \frac{4}{7}(b_2^{lm})^2$ and $u_2 = u_2^n + \frac{4}{7}b_0^{lm}b_2^{lm} - \frac{1}{7}\varepsilon_0^{nl}\varepsilon_2^{nl}$, according to eq (AII.6). The temperature dependence of the optimal values of these two parameters of the potential for the *p*-dcb molecule, obtained from the experimentally determined solute order parameters in Figure 3 is shown in Figure 6. The two parameters do not appear to change discontinuously across the N-$N_X$ phase transition; however they show different trends of variation with temperature in the two phases and this could be attributed to the onset of the quadratic contributions to $u_0$ and $u_2$ in the above expressions on entering the $N_X$ phase.

**5.2.2. The trichlorobenzene (tcb)** molecule has a threefold symmetry axis normal to the ring plane, identified as the *x* molecular axis (see Figure 2c). The *z* axis is taken along any intramolecular Cl-H diameter and *y* is then parallel to the corresponding Cl-Cl direction. Therefore, $\langle S_{xx}^H \rangle$ is the only surviving independent order parameter ($\langle S_{yy}^H \rangle = \langle S_{zz}^H \rangle = -\langle S_{xx}^H \rangle/2$) that determines the spectral signature and we have for the three equivalent proton pairs of the tcb molecule

$$(\nu/\nu^0)^{tcb} = -\frac{1}{2}\langle S_{xx}^H \rangle \quad . \tag{35}$$

The order parameter of tcb can thus be obtained from the single dipolar coupling measured by NMR. Furthermore, if the plausible assumption is made that the order parameter $\langle S_{xx}^H \rangle$ is nearly the same for the trichlorobenzene and the *p*-dichlorobenzene solutes, at the same solvent temperature, then the combination of eq (35) with the first two relations in eq (32) implies that the following linear relation among the tcb and *p*-dcb experimental dipolar couplings should hold:

$$D^{tcb} = \kappa_{ortho} D_{ortho}^{p-dcb} + \kappa_{meta} D_{meta}^{p-dcb} \quad , \tag{36}$$

with $\kappa_{ortho} = \frac{1}{2}(r_{ortho}^{p-dcb}/r^{tcb})^3$ and $\kappa_{meta} = \frac{1}{2}(r_{meta}^{p-dcb}/r^{tcb})^3$ representing temperature independent geometrical constants. As shown in Figure 8, this linear relation appears to hold rather well for the measured spectra of the two molecules, with a slope of 0.096 and intercept 0.507. Again these values are close but not coincident with the theoretical values obtained from quantum chemistry calculations[27] of the inter-proton distances yielding $\kappa_{ortho} = 0.0914$ and $\kappa_{meta} = 0.50$.

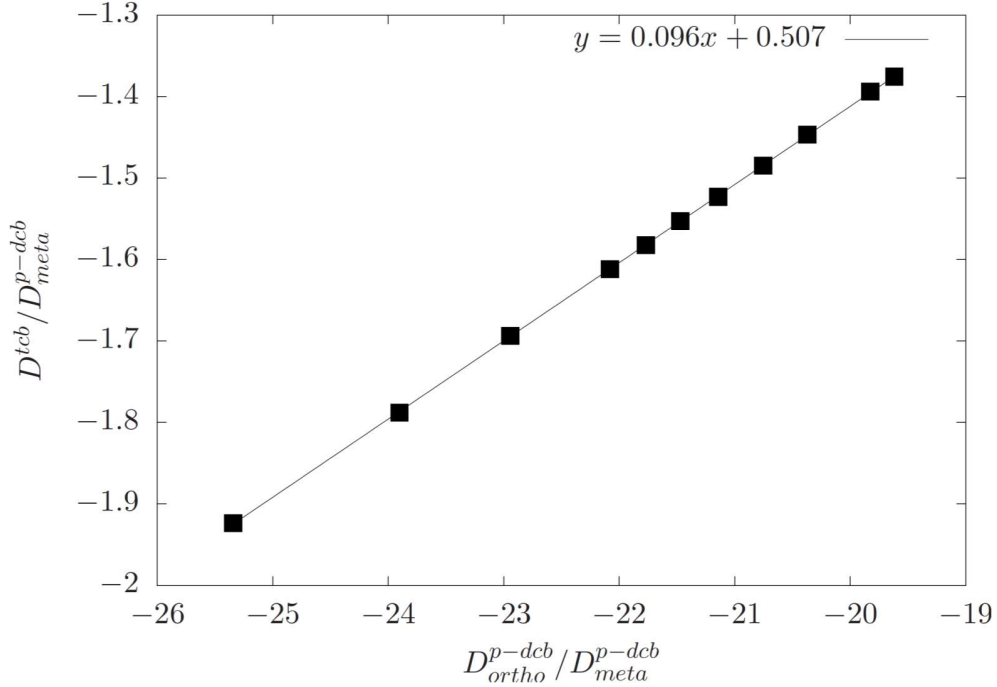

**Figure 8.** Plot of the ratios of measured[22] dipolar coupling of the *p*-dcb and tcb solutes in the N and $N_X$ phases of a CBC9CB/5C mixture showing a linear relation, in agreement with eq(36) which is based on the near equality of the $\langle S_{xx}^H \rangle$ order parameters of the two solutes.

The threefold symmetry of the tcb molecule about the *x*-axis implies, in addition to the vanishing of the polar and chiral contributions, the following relations between the basic parameters of the remaining part of the potential of mean torque in eq (28), $u_0^n = -\frac{2}{3}u_2^n$; $b_0^{lm} = -b_2^{lm}$; $\varepsilon_0^{nl} = -\frac{1}{2}\varepsilon_2^{nl}$ leading to the relation $u_0 = -\frac{2}{3}u_2$ for the parameters of the rotationally averaged potential in eq (34), which therefore in the case of tcb reduces to a single parameter potential

$$\overline{V}(\theta,\psi)\big|_{tcb} = 2u_2\left(\sin^2\theta\cos^2\psi - \frac{1}{3}\right) \quad . \tag{37}$$

We have used the measured dipolar couplings of tcb in the N and $N_X$ phases of CBC9CB/5C mixtures[22] to determine the single parameter $u_2$ in the above equation. The temperature dependence of the optimal values of this parameter, obtained from the calculated values of $\langle S_{xx}^H \rangle$ in eq (35), is shown in Figure 6. The near coincidence of the optimal values of $\langle S_{xx}^H \rangle$ for the *p*-dcb and tcb solutes is reflected on the near equality of the pertinent combinations of the coefficients of the potentials in eqs (34) and (37) for the two solutes, namely $\left[u_2 - \frac{u_0}{2}\right]_{p-dcb} \approx \left[u_2 - \frac{u_0}{2}\right]_{tcb} = \frac{4}{3}[u_2]_{tcb}$, which is closely satisfied by the optimal values plotted in Figure 6. The combination $u_0 + \frac{2}{3}u_2$, associated with the effect of molecular biaxiality on the ordering of the two axes that define the solute's molecular plane (Figures 2b and c), shows substantial values in the plots of Figure 6 for $p$-dcb, in accord with the lack of equivalence between the $y$ and $z$ molecular axes, and vanishes by symmetry for the tcb solute.

**5.2.2. The anthracene** molecule, see Figure 2d, has identical symmetries with the $p$-dcb molecule. Accordingly, its orientational ordering as a solute in the N and $N_X$ phases is described by the two order parameters that appear in eq (32) and the respective potential of mean torque has the parameterization of the form given in eq (34). The measured deuterium quadrupolar NMR spectrum of anthracene as a solute in CB7CB[23] provides in either of the nematic phases two groups of distinct splittings, from which the two order parameters and the effective potential parameters $u_0, u_2$ can be determined. The temperature dependence of the respective values is shown in Figures 3 and 6. Anthracene and $p$-dcb, having the same molecular symmetry and nearly the same lateral dimensions, present in Figure 3 nearly equal values for the order parameter $\langle S_{xx}^H \rangle$ while the elongation of the anthracene molecule is reflected on somewhat larger values of $\langle S_{zz}^H \rangle$ compared to those of $p$-dcb. Interestingly, $\langle S_{zz}^H \rangle$ for anthracene shows a jump across the N-$N_X$ phase transition, unlike its apparently continuous evolution in the case of the $p$-dcb solute which, however, is dissolved in a different solvent. Analogous trends and relative magnitudes are observed in Figure 6 for the parameters of the potentials describing the ordering of the two solutes: $u_0$ showing larger values for anthracene and a discontinuity at the phase transition and similar values for the $u_2$ parameters of the two solutes.

Beyond demonstrating the adequacy of the proposed potential of mean torque to fully account for the measured NMR spectra of a variety of small rigid solutes in the N and $N_X$ phase, the collective plotting of the parameters $u_0, u_2$ and $|u_1|$ in Figure 6 marks out the internal consistency of the potential as it shows that the optimal values of these parameters and the trends of their

variation are in physical accord with the symmetry, molecular axis assignment and variation in the size and shape of the solutes.

## 6. Discussion.

Enantiotopic discrimination is observed on the NMR spectra of certain types of prochiral solutes in the N* and the $N_X$ phases and is in both cases attributed to the existence of some chiral asymmetry that can be "read" by the prochiral elements of the probe solutes. However, the local structure and symmetry of the two nematic phases differ significantly and the question is raised as to how these differences are reflected on the mechanisms that give rise to such chiral asymmetry. The formulation of the potential of mean torque on the basis of solvent-solute molecular interactions, local phase symmetry and solute molecular symmetry, as done here for the $N_X$ phase and in ref. [21] for the N* phase, allows the identification of the possible mechanisms and the quantification of their respective contributions to the generation of the overall chiral asymmetry in the environment of the solute molecules. The leading contributions to the potential of mean torque governing the orientational ordering of solutes in the $N_X$ phase indicate the possibility of three distinct mechanisms for the generation of chiral asymmetry in the solvent medium. These three mechanisms, the relevance and significance of which differs between the $N_X$ and N* phases, are summarized below:

(i) Local structural chirality, produced by the polarity of the local ordering in the direction ($\hat{m}$) transverse to the helix axis, combined with the eccentricity/tilt of the ordering in the plane perpendicular to the polar director $\hat{m}$. As a result of this mechanism, the local environment felt by the prochiral molecular fragments of the solute discriminates between enantiotopic elements. Obviously, to couple to this mechanism, the solute molecules should be polar. Local structural chirality emerges as the primary mechanism of enantiotopic discrimination for prochiral solutes of small molecular dimensions in the $N_X$ phase and is applicable only for molecules of the $C_s$ and $C_{2v}$ point symmetries. This mechanism is clearly not applicable in the N* phase due to the lack of local polar ordering therein.

(ii) Chirality of the solvent molecules generates, through direct interactions with the solvent molecules, an environment that discriminates between enantiotopic elements of the latter. This mechanism has been shown to underlie entirely enantiotopic discrimination phenomena in the N* phase[21] and to be applicable for solute molecules belonging to the point group symmetries $C_s$, $C_{2v}$, $D_{2d}$ and $S_4$. An analogous mechanism is also shown here to be in principle operative in the $N_X$ phase as well, albeit under different local phase symmetry. Molecular chirality results in this case from the possible loss of strict statistical achirality of the solvent molecules under the tight twisting of the polar director $\hat{m}$. It is, however, expected to contribute marginally to the enantiotopic discrimination as the extent of the induced statistical chirality of the dimer molecules in the $N_X$ phase is estimated to be rather limited.[19] Moreover, the $N_X$ phase can in principle be formed by perfectly rigid achiral molecules that would, of course, maintain their strict achirality in the twisted domains of the $N_X$ phase. A quantitative estimate of the contribution of molecular chirality to the enantiotopic discrimination in the $N_X$ phase can be obtained from NMR experiments using apolar prochiral solutes which show measurable enantiotopic discrimination in the N* phase. For example, spiropentane,[15,16] being of $D_{2d}$ symmetry would not couple to the local polar ordering and therefore, ignoring molecular size effects (see below), any enantiotopic discrimination it might show in the $N_X$ phase would be attributable to the mechanism of molecular chirality.

(iii) Chirality resulting from the spatial modulation of the directors. The previous two mechanisms are the only applicable ones to small solutes, i.e. solutes whose spatial extension is negligible compared to the pitch of the helical twisting of the phase director. For relatively extended solutes, however, the possibility of different parts of the same solute molecule sampling regions of the solvent medium in which the director is oriented differently, may in general influence their NMR spectra. If the spatial modulation is chiral, such influence could lead to discrimination between enantiotopic elements of a prochiral solute. The spontaneous spatial modulations in both the N* and $N_X$ phases consist of helical twisting and are therefore chiral. There are, however, important differences between the two phases: The twisting director in the N* phase is the apolar nematic director $\hat{n}$ whereas the twisting director in the $N_X$ phase is the polar director $\hat{m}$. Both directors twist at right angles to the helix axis, however the pitch in the $N_X$ phase is on the length scale of the dimer molecular dimensions whereas the N* pitch is longer by two or more orders of magnitude. Accordingly, solutes of molecular dimensions of a few *nm* are considered as spatially extend in the $N_X$ phase. In contrast, the dimensions of the same solutes would be treated as negligible in the N* phase since their different parts would sample a region of the solute medium where the orientation of the director is practically uniform. Therefore, the contribution of spatial modulations would be negligible for such solutes in the N* phase. In fact spatial modulation of the director in the N* phase is removed altogether in NMR experiments on samples where the helix is unwound under the action of the spectrometer magnetic field; yet enantiotopic discrimination is clearly observed on the spectra. On the other hand, the tight twisting of the $\hat{m}$ director in the $N_X$ phase renders the spatial modulation a potentially significant mechanism of enantiotopic discrimination for solutes of comparable dimensions with the solvent dimer molecules. The effects of this mechanism depend directly on the solute molecular dimensions relative to the helical pitch and combine with the effects of mechanisms (i) and (ii) above. Last, in all the known NMR experiments in the $N_X$ phase, the helical pitch is practically unaffected by the spectrometer field.

To our knowledge, the only other attempt in the literature to date at a molecular interpretation of enantiotopic discrimination in the $N_X$ phase is formulated entirely on the basis of director modulations[28]. There, however, the twisting director is not the polar director $\hat{m}$, as described in (iii) above, but a "nematic director" $\hat{n}$ (i.e a local axis of apolar uniaxial orientational ordering) which forms a constant angle $\theta_0$ with the helix axis, according to the usual picture of the twist-bend model. Such formulation apparently corresponds to a generalization of the mechanism of director modulations in the N* phase, summarized in (iii) above, based on the idea the N* phase can be viewed as a special case of the twist-bend nematic when $\theta_0$ is 90°.[12] This generalization clearly disregards the significance of the two or more orders of magnitude difference between the pitch of the N* and $N_X$ phase and ignores any local polar ordering in the latter phase. Such approach, however, could not account, using any reasonable value for the pitch, for the enantiotopic discrimination exhibited by small solute molecules in the $N_X$ phase. Furthermore, it would obviously yield a null result in place of the finite values that are actually measured by NMR for the enantiotopic discrimination exhibited by various prochiral solutes in N* solvents with magnetically unwound helix.

To obtain the order parameter profiles of the prochiral solute acenaphthene in Figure 3 and the respective values of the potential parameters in Figure 6, the spectra were analyzed assuming that, on the timescale of the NMR measurement, a solute molecule remains within a domain of given handedness in the $N_X$ phase. This assumption of slow migration of molecules from one domain to another appears to be fully supported by the analysis of the NMR spectra obtained, both parallel and perpendicular to the magnetic field, from labeled molecules of CB9CB,[13] at least at sufficiently low temperatures in the $N_X$ phase. Such support, however, does not guarantee slow mobility for small solute molecules as well, since these could show considerably higher mobility than the solvent dimer molecules. On the other hand, the fact that enantiotopic discrimination is indeed observed on the spectra of acenaphthene indicates that these molecules do not migrate very rapidly between domains of opposite twisting sense, although it cannot be excluded that the molecules might on average spend a non-negligible fraction of the NMR measurement time residing in domains of opposite chirality. In such case, the measured spectral signature differences associated with enantiotopic discrimination would appear reduced by that fraction and therefore the apparent values of the order parameter $\langle S_{xy}^{H} \rangle$ in eq (23) would be proportionately reduced with respect to the actual values. Accordingly, the values plotted in Figure 3 for $\langle S_{xy}^{H} \rangle$ obtained from eq (23), should be considered as lower bounds that correspond to negligible inter-domain migration and negligible molecular deformation of the solute molecules. Extending these considerations, one could interpret the complete lack of enantiotopic discrimination in the high temperature N phase in two ways, both of which are acceptable on the basis of the NMR spectral evidence alone: One interpretation would be that the high temperature nematic phase is a usual apolar uniaxial nematic. The other interpretation would be that the high temperature nematic phase has the same local symmetries as the $N_X$ phase only the size of the domains is much smaller and consequently a solute molecule would reside in a large number of such domains of opposite twisting sense during the NMR time window, leading to the complete vanishing of the difference between the frequencies of the enantiotopic sites. Such interpretation cannot be ruled out by the NMR data nor by the proposed potential of mean torque, since all the terms that give rise to the order parameter $\langle S_{xy}^{n_h} \rangle$, which in turn generates the enantiotopic discrimination according to eq (19), invert their sign on moving between domains of opposite twisting sense. Furthermore, examples are known for the possibility of phase transitions between nematic phases that differ only in the size of the molecular clusters into which the constituent molecules are organized.[29–33]

Quantitative estimates of the size of the twisted domains in the $N_X$ phase can be obtained from the NMR spectra in conjunction with information on molecular diffusion. The experimental NMR data on acenaphthene were originally concluded[12] to be consistent with either the formation of a uniformly handed phase or with slow inter-domain diffusion. It was later shown[13] that slow diffusion, combined with measured NMR spectra perpendicular to the magnetic field, would eliminate the twist-bend interpretation of the $N_X$ phase, as the measured spectra were found to differ qualitatively from those that would be obtained from a phase with a heliconically modulated nematic director. It was also shown that the opposite possibility, of fast intra-domain diffusion, would simply mean that the NMR measurements are not directly sensitive to any sort of twisting within a domain. Recently, values of the diffusion coefficients in CB7CB were reported[34] to be on the order of $3 \times 10^{-11}$ m$^2$s$^{-1}$ for diffusion along the helix axis and roughly one third of that value for diffusion in the perpendicular direction. The respective values for smaller molecules, such as acenaphthene, could be reasonably expected to be larger by as much as an order of magnitude.[35] This would place the average spatial displacement of a small solute molecule in the N$_X$ phase during the NMR time window on the order of $10^{-5}$ m. Such displacements would of course average out the details of any modulation along the twisting axis, and, given the clear detection of enantiotopic discrimination, would also place a lower bound on the dimensions of the twisted domains in the N$_X$ phase.

## 7. Conclusions.

Using a systematic approach for the formulation of the potential of mean torque in the N$_X$, N* and N phases it is possible to describe consistently the orientational ordering of small rigid solutes dissolved in these phases and to identify, in terms of molecular interactions, the similarities and differences of the underlying ordering mechanisms. Chiral asymmetry of the orientational ordering in the N* and N$_X$ phases leads to enantiotopic discrimination phenomena but through distinctly different mechanisms: Whilst molecular chirality of the solvent molecules accounts fully for enantiotopic discrimination presented exclusively by solutes of point group symmetries $C_s$, $C_{2v}$, $D_{2d}$ and $S_4$ in the N* phase, it plays a marginal role in the N$_X$ phase. Structural chirality, i.e. the simultaneous presence of polar ordering and eccentricity, is found to be the primary cause of enantiotopic discrimination for small rigid prochiral solutes in the N$_X$ phase and is totally absent in the N* phase. Director modulations, either in the form of nematic director twisting for the N* phase or in the form of polar director twisting in the N$_X$ phase, contribute significantly to enantiotopic discrimination only for extended solutes in the N$_X$ phase.

As structural chirality refers only to prochiral solutes of the $C_s$ and $C_{2v}$ symmetries, assessment of its contribution to the enantiotopic discrimination relative to the -expected marginal- contribution of molecular chirality in the N$_X$ phase can be provided by NMR measurements on small rigid prochiral solutes of the $D_{2d}$ or $S_4$ symmetry, which are known to exhibit such discrimination in the N* phase

For the small rigid solutes considered in this study, the presence of enantiotopic discrimination indicates limited diffusion between domains of opposite twisting sense during the NMR measurement time. Accordingly, estimates based on values reported in the literature for the diffusion coefficients would place the lower bound of the spatial extent of the twisted domains in the $N_X$ phase on the order of 10 μm.

**Appendix I. Accounting for the spatial extension of the solute molecules.**

Noting that the tensor contributions to $V(\omega, Z)$ of eq (10) include up to quadratic terms in the twisting axes $\hat{m}$ and $\hat{l}_h$, this potential can be expressed in terms of the Euler angles $\theta, \psi, \varphi$ in the alternative form

$$V(\omega, Z) = \sum_{\mu=-2}^{2} U^{(\mu)}(\theta, \psi) e^{i\mu\tilde{\varphi}} , \qquad (AI.1)$$

where the angle $\tilde{\varphi} = \varphi - k(Z - Z_0)$ is introduced in the paragraph preceding eq (30) and the five $U^{(\mu)}$ components ($\mu = 0, \pm 1, \pm 2$) of the potential, which contain the $\theta$ and $\psi$ dependence of the latter, satisfy the condition imposed by the reality of the total potential $V$, namely $U^{(\mu)} = \left(U^{(-\mu)}\right)^*$.

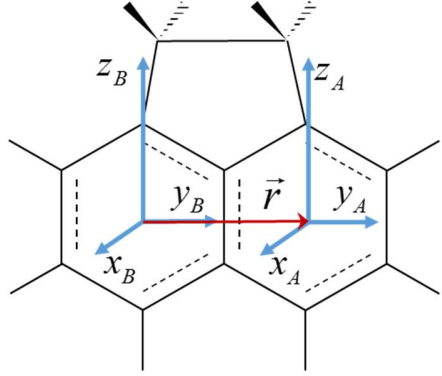

**Figure A1.** Partitioning of a solute molecule into two submolecular halves (*A* and *B*), with the origins of the respective submolecular axis frames separated by the intra molecular vector $\vec{r}$.

Due to the tight twisting of the polar director $\hat{m}$, distant parts of a relatively extended solute molecule will sample different directions of $\hat{m}$, or, equivalently, of the angle $\tilde{\varphi}$ of eq (AI.1). To illustrate how this would affect the overall potential of the entire molecule we consider the example of a rigid achiral solute molecule that is treated as two distinct parts *A* and *B* and attach submolecular axis frames to each one of these parts (see Figure A1); the frames are taken to have their respective axes parallel to each-other and to the axes of the "central" molecular frame. Suppose that the origins of these axes are located at positions $\vec{r}_A = -\vec{r}$ and $\vec{r}_B = \vec{r}$. Then, expressing the total potential of mean torque $V$ of the solute molecule as a sum of two terms, $V_A, V_B$, we have, in analogy with eq (AI.1)

$$V_{A(B)}(\omega, Z) = \sum_{\mu=-2}^{2} U_{A(B)}^{(\mu)}(\theta, \psi) e^{i\mu \tilde{\varphi}_{A(B)}} \quad , \tag{AI.2}$$

with $\tilde{\varphi}_A = \tilde{\varphi} + k(\hat{n}_h \cdot \vec{r})$, $\tilde{\varphi}_B = \tilde{\varphi} - k(\hat{n}_h \cdot \vec{r})$ and $\tilde{\varphi}$ referring to the center of the solute molecule; therefore the total potential is obtained as

$$V(\omega; Z) = V_A(\omega; Z_A) + V_B(\omega; Z_B) = \sum_{\mu=-2}^{2} U_{comb}^{(\mu)}(\theta, \psi) e^{i\mu \tilde{\varphi}} \quad . \tag{AI.3}$$

The total potential in the above equation is of the same general form with eq (AI.1), except that the components $U^{(\mu)}$ are here replaced by the combined components $U_{comb}^{(\mu)} = U_A^{(\mu)} e^{i\mu k(\hat{n}_h \cdot \vec{r})} + U_B^{(\mu)} e^{-i\mu k(\hat{n}_h \cdot \vec{r})}$ which show explicit dependence on the helical twisting wavenumber $k$. Naturally, this dependence becomes weaker for molecules whose spatial extension, represented by the intramolecular vector $\vec{r}$, becomes small compared to $1/k$, in which case eq (AI.3) tends to eq (AI.1). It should be noted that the subdivision of the molecule into fragments generally introduces asymmetries that are not present in the entire molecule. Thus for example, the sub-molecular fragments of an apolar achiral solute molecule could be polar or chiral.

**Appendix II.**

**II.1.** The leading rank terms of the complete potential of mean torque for a small rigid solute molecule of $C_{2v}$ symmetry in the $N_X$ phase.

To describe the solute molecule's orientation relative to the right-handed macroscopic $X, Y, Z$ frame we use the two Euler angles $\theta, \psi$ and the angle $\tilde{\varphi}$ introduced in section 5.1. For a solute molecule of $C_{2v}$ symmetry, with the molecular axes $x, y, z$ defined as in Figure 2 for the acenaphthene molecule, we have, according to the symmetry implications in section 4.2, the following contributions to the potential of mean torque $V(\omega, Z) = V(\theta, \psi; \tilde{\varphi})$:

The polar contribution

$$V^{(1)}(\theta, \tilde{\varphi}) = u_{polar} \sin\theta \sin\tilde{\varphi} \quad , \tag{AII.1}$$

with the parameter $u_{polar}$ conveying the coupling of the polar direction of the solute, i.e. the $z$ molecular axis, to the polar director $\hat{m}$ of the $N_X$ phase.

The second rank symmetric contribution

$$V^{(2)}(\theta,\psi;\tilde{\varphi}) = u_0^n \left(\frac{3}{2}\cos^2\theta - \frac{1}{2}\right) + u_2^n \sin^2\theta \cos 2\psi +$$
$$+ b_0^{lm} \sin^2\theta \cos 2\tilde{\varphi} + b_2^{lm} \left((\cos^2\theta + 1)\cos 2\psi \cos 2\tilde{\varphi} - 2\cos\theta \sin 2\psi \sin 2\tilde{\varphi}\right) + \quad , \quad \text{(AII.2)}$$
$$+ \varepsilon_0^{nl} \sin 2\theta \cos\tilde{\varphi} + \varepsilon_2^{nl} \left(\sin\theta \sin 2\psi \sin\tilde{\varphi} - \frac{1}{2}\sin 2\theta \cos 2\psi \cos\tilde{\varphi}\right)$$

where the coefficients are given in terms of the solvent order parameters and the solvent-solute molecular coupling parameters of eq (8) according to:

$$u_0^n = \langle g_{zz}^{(2)}\rangle' - \frac{1}{2}\left(\langle g_{xx}^{(2)}\rangle' + \langle g_{yy}^{(2)}\rangle'\right); \qquad u_2^n = \frac{3}{4}\left(\langle g_{xx}^{(2)}\rangle' - \langle g_{yy}^{(2)}\rangle'\right)$$
$$b_0^{lm} = \frac{3}{2}\left(\langle \Delta_{zz}^{(2)}\rangle' - \frac{1}{2}\left(\langle \Delta_{xx}^{(2)}\rangle' + \langle \Delta_{yy}^{(2)}\rangle'\right)\right); \qquad b_2^{lm} = \frac{3}{4}\left(\langle \Delta_{xx}^{(2)}\rangle' - \langle \Delta_{yy}^{(2)}\rangle'\right) \quad , \quad \text{(AII.2')}$$
$$\varepsilon_0^{nl} = \langle \Phi_{zz}^{(2)}\rangle' - \frac{1}{2}\left(\langle \Phi_{xx}^{(2)}\rangle' + \langle \Phi_{yy}^{(2)}\rangle'\right); \qquad \varepsilon_2^{nl} = \langle \Phi_{xx}^{(2)}\rangle' - \langle \Phi_{yy}^{(2)}\rangle'$$

Last, the presence of induced molecular chirality on the statistically achiral dimer molecules forming the $N_X$ phase would introduce chiral contributions of the vector-pseudovector form given in eq (10). This part of the potential of mean torque has the following expression in terms of the angles $\theta, \psi$ and $\tilde{\varphi}$:

$$V^*(\theta,\psi;\tilde{\varphi}) = c^{n*} \sin^2\theta \sin 2\psi + c^{lm*}\left[(\cos^2\theta + 1)\sin 2\psi \cos 2\tilde{\varphi} + 2\cos\theta \cos 2\psi \sin 2\tilde{\varphi}\right] +$$
$$+ c_+^{nl*}\left(\sin 2\theta \sin 2\psi \cos\tilde{\varphi} + 2\sin\theta \cos 2\psi \sin\tilde{\varphi}\right) + c_-^{nl*} \sin\theta \sin\tilde{\varphi} \quad \text{(AII.3)}$$

with the coefficients representing the following combinations of solvent order parameters and chiral solvent- solute molecular coupling parameters:

$$c^{n*} = 4\left[g_{x\times z,x;a'\neq b',c'}^{(p)} - g_{y\times z,y;a'\neq b',c'}^{(p)}\right]\left\langle\frac{3}{2}\left((\hat{a}'\times\hat{b}')\cdot\hat{n}_h\right)(\hat{c}'\cdot\hat{n}_h) - \frac{1}{2}\varepsilon_{a',b',c'}\right\rangle'$$
$$c^{lm*} = \frac{9}{4}\left[g_{x\times z,x;a'\neq b',c'}^{(p)} - g_{y\times z,y;a'\neq b',c'}^{(p)}\right]\left\langle\left((\hat{a}'\times\hat{b}')\cdot\hat{l}_h\right)(\hat{c}'\cdot\hat{l}_h) - \left((\hat{a}'\times\hat{b}')\cdot\hat{m}\right)(\hat{c}'\cdot\hat{m})\right\rangle'$$
$$c_+^{nl*} = \frac{1}{2}\left(g_{y\times z,y;a'\neq b',c'}^{(p)} - g_{x\times z,x;a'\neq b',c'}^{(p)}\right)\left[\left\langle\left((\hat{a}'\times\hat{b}')\cdot\hat{l}_h\right)(\hat{c}'\cdot\hat{n}_h)\right\rangle' + \left\langle\left((\hat{a}'\times\hat{b}')\cdot\hat{n}_h\right)(\hat{c}'\cdot\hat{l}_h)\right\rangle'\right] \quad \text{(AII.3')}$$
$$c_-^{nl*} = \left(g_{y\times z,y;a'\neq b',c'}^{(p)} + g_{x\times z,x;a'\neq b',c'}^{(p)}\right)\left(\left\langle\left((\hat{a}'\times\hat{b}')\cdot\hat{l}_h\right)(\hat{c}'\cdot\hat{n}_h)\right\rangle' - \left\langle\left((\hat{a}'\times\hat{b}')\cdot\hat{n}_h\right)(\hat{c}'\cdot\hat{l}_h)\right\rangle'\right)$$

**II.2. The effective potential for the solute $\tilde{\varphi}$ -averaged orientational distribution.**

In using the total potential of mean torque $V(\theta,\psi;\tilde{\varphi}) = V^{(1)} + V^{(2)} + V^*$ to evaluate the solute order parameters appearing in eq (19), namely

$$\langle S_{zz}^{n_h} \rangle = \langle \frac{3}{2}\cos^2\theta - \frac{1}{2} \rangle$$

$$\langle S_{xx}^{n_h} \rangle - \langle S_{yy}^{n_h} \rangle = \frac{3}{2}\langle \sin^2\theta \cos 2\psi \rangle \quad , \tag{AII.4}$$

$$\langle S_{xy}^{n_h} \rangle = -\frac{3}{4}\langle \sin^2\theta \sin 2\psi \rangle$$

we note that these do not depend on translations along, or rotations about, the Z axis and therefore they are determined by the probability distributions of just the $\theta$ and $\psi$ angles. This $\tilde{\varphi}$-averaged distribution is obtained from the effective potential $\bar{V}(\theta,\psi)$, defined according to $e^{\bar{V}(\theta,\psi)} = (1/2\pi)\int_0^{2\pi} e^{V(\theta,\psi;\tilde{\varphi})} d\tilde{\varphi}$. To get an analytic expression for the leading rank contributions of the effective potential $\bar{V}(\theta,\psi)$ in terms of the parameters specifying the potential of mean torque $V(\theta,\psi;\tilde{\varphi})$ in eqs(AII.1-3) we express the latter as a sum, $V(\theta,\psi;\tilde{\varphi}) = W(\theta,\psi) + W'(\theta,\psi;\tilde{\varphi})$, of two parts, one, $W(\theta,\psi)$, that is independent of rotations about the helix axis Z or translations along that axis, and a second part, $W'(\theta,\psi;\tilde{\varphi})$, that depends on such rotations/translations and will therefore involve the $\tilde{\varphi}$ angle. Next, we make use of $|W'(\theta,\psi;\tilde{\varphi})| < 1$ and $\int_0^{2\pi} W'(\theta,\psi;\tilde{\varphi})d\tilde{\varphi} = 0$ to approximate $(1/2\pi)\int_0^{2\pi} e^{W'(\theta,\psi;\tilde{\varphi})} d\tilde{\varphi} \approx 1 + (1/4\pi)\int_0^{2\pi} [W'(\theta,\psi;\tilde{\varphi})]^2 d\tilde{\varphi}$. It then follows that $\bar{V}(\theta,\psi) \approx W(\theta,\psi) + (1/4\pi)\int_0^{2\pi} [W'(\theta,\psi;\tilde{\varphi})]^2 d\tilde{\varphi}$ and, aside from trivial constant terms, the leading tensor rank expression for the effective potential is

$$\bar{V}(\theta,\psi) = u_0\left(\frac{3}{2}\cos^2\theta - \frac{1}{2}\right) + u_2 \sin^2\theta \cos 2\psi + u_1 \sin^2\theta \sin 2\psi + 0(rank\, l \geq 4\, terms) \quad . \tag{AII.5}$$

The coefficients in this expression are related to the coefficients describing the various parts of the initial potential of mean torque in eqs(AII.1-3) as follows:

$$u_0 = u_0^n + \frac{2}{21}(\varepsilon_0^{nl})^2 - \frac{1}{14}(\varepsilon_2^{nl})^2 - \frac{2}{7}(c_+^{nl*})^2 - \frac{1}{6}(u_{polar} + c_-^{nl*})^2 + \frac{4}{7}\left((b_2^{lm})^2 + (c^{lm*})^2\right) - \frac{4}{21}(b_0^{lm})^2$$

$$u_2 = u_2^n + c_+^{nl*}(u_{polar} + c_-^{nl*}) + \frac{4}{7}b_0^{lm}b_2^{lm} - \frac{1}{7}\varepsilon_0^{nl}\varepsilon_2^{nl} \qquad (AII.6)$$

$$u_1 = c^{n*} + \frac{1}{2}\varepsilon_2^{nl}(u_{polar} + c_-^{nl*}) + \frac{4}{7}b_0^{lm}c^{lm*}$$